\documentclass[10pt, conference, letterpaper]{IEEEtran}
\IEEEoverridecommandlockouts
\usepackage{cite}
\usepackage{amsthm}
\usepackage{amsmath,amssymb,amsfonts}
\usepackage{algorithm}
\usepackage{algorithmicx}
\usepackage{algpseudocode}
\usepackage{amsmath}
\usepackage{mathtools}
\usepackage{graphicx}
\usepackage{textcomp}
\usepackage{xcolor}
\usepackage{bm}
\usepackage{fmtcount}
\usepackage{graphicx,epstopdf,psfrag,url,cite,xcolor,multirow,float}
\usepackage[font=footnotesize]{subfig}

\graphicspath{{figures/}}
\def\BibTeX{{\rm B\kern-.05em{\sc i\kern-.025em b}\kern-.08em
    T\kern-.1667em\lower.7ex\hbox{E}\kern-.125emX}}
\begin{document}
\title{Digital Twin Enabled Simultaneous Learning and Modeling for UAV-assisted Secure Communications with Eavesdropping Attacks}
\author{Jieting Yuan, Songhan Zhao, Ye Xue, Yu Zhao, Bo Gu, and Shimin Gong

\thanks{Jieting Yuan, Songhan Zhao, Ye Xue, Bo Gu, and Shimin Gong  are with the School of Intelligent Systems Engineering, Sun Yat-sen University, China (e-mail: \{yuanjt3, zhaosh55\}@mail2.sysu.edu.cn; \{xuey57, gubo, gongshm5\}@mail.sysu.edu.cn). Yu Zhao is with the Department of Equipment Management and Unmanned Aerial Vehicle Engineering, Air Force Engineering University, China (e-mail: zhaoyuair@163.com).

}
}

\maketitle

\begin{abstract}
This paper focuses on secure communications in UAV-assisted wireless networks, which comprise multiple legitimate UAVs (LE-UAVs) and an intelligent eavesdropping UAV (EA-UAV).  The intelligent EA-UAV can observe the LE-UAVs’ transmission strategies and adaptively adjust its trajectory to maximize information interception. To counter this threat, we propose a mode-switching scheme that enables LE-UAVs to dynamically switch between the data transmission and jamming modes, thereby balancing data collection efficiency and communication security. However, acquiring full global network state information for LE-UAVs' decision-making incurs significant overhead, as the network state is highly dynamic and time-varying. To address this challenge, we propose a digital twin-enabled simultaneous learning and modeling (DT-SLAM) framework that allows LE-UAVs to learn policies efficiently within the DT, thereby avoiding frequent interactions with the real environment. To capture the competitive relationship between the EA-UAV and the LE-UAVs, we model their interactions as a multi-stage Stackelberg game and jointly optimize the GUs' transmission control, UAVs' trajectory planning, mode selection, and network formation to maximize overall secure throughput. Considering potential model mismatch between the DT and the real environment, we propose a robust proximal policy optimization (RPPO) algorithm that encourages LE-UAVs to explore service regions with higher uncertainty. Numerical results demonstrate that the proposed DT-SLAM framework effectively supports the learning process. Meanwhile, the RPPO algorithm converges about $12\%$ faster and the secure throughput can be increased by $8.6\%$ compared to benchmark methods.
\end{abstract}

\begin{IEEEkeywords}
Secure communications, digital twin, Stackelberg game, PPO
\end{IEEEkeywords}

\section{Introduction}
Unmanned aerial vehicles (UAVs) have emerged as mobile aerial communication platforms with significant potential for environmental monitoring, emergency communications, disaster rescure, and internet of things (IoT) networks~\cite{4mozaffari2019tutorial}. By adjusting their positions dynamically, UAVs are capabe of reducing communication distances and improving channel conditions. Multi-UAV networks with UAV-to-UAV (U2U) connections can expand coverage and improve performance in large-scale distributed communication systems~\cite{6gong2023bayesian}. However, UAV-assisted wireless channels are highly vulnerable to eavesdropping attack due to their broadcasting nature and low-altitude transmissions~\cite{7pandey2022security}. Intelligent eavesdropping UAVs (EA-UAVs) can exploit this vulnerability by optimizing their flight paths in real time to maximize the interception of confidential information transmitted among legitimate UAVs (LE-UAVs). Although cryptographic methods can encrypt data prior to transmission~\cite{8aissaoui2023survey}, they typically require substantial computational resources. This renders them impractical for real-time UAV-assisted IoT applications, where processing capability is limited and data traffic is bursty~\cite{9sun2019physical}. In contrast, physical-layer security (PLS) exploits UAVs' mobility through joint trajectory planning and transmission control. Beyond passive reactions, LE-UAVs' active jamming can be employed to degrade the eavesdropper's channel conditions by beamforming artificial interference signals, thereby reducing  EA-UAVs' ability to decode transmitted information.

\subsection{Motivations and Challenges}
Existing UAV-assisted secure communication systems often assume simplified eavesdropper behaviors, such as static positions or predictable trajectories~\cite{16dan2024proactive,17wen2021securing,18zhang2024one}. However, when EA-UAVs can adaptively adjust their  eavesdropping strategies, these assumptions lead to ineffective security measures in practice. Meanwhile, most existing works assign fixed roles to LE-UAVs, i.e., some LE-UAVs are dedicated to transmission and another set of LE-UAVs are designed as active jammers~\cite{13liu2023jamming,15ye2023robust}. Such static configurations are also vulnerable to more intelligent EA-UAVs that can deceive the LE-UAVs and adapt their eavesdropping strategies. For example, the EA-UAV can stay far away from jamming LE-UAVs and follow the data-transmitting ones. As such, the jamming LE-UAVs' interfering signals become ineffective, wasting energy and spectral resources. Therefore, it is essential to develop an adaptive strategy for LE-UAVs that dynamically optimizes their mode selection, trajectory planning and resource allocation in response to EA-UAVs'  eavesdropping behaviors.


Moreover, the traffic demands of ground users (GUs) in practical networks are often highly dynamic, rendering pre-defined
control strategies ineffective. Though deep reinforcement learning (DRL) can help LE-UAVs to learn optimal policies through continuous interactions with the network environment, it typically suffers from slow convergence and poor sample efficiency due to costly action exploration in the real environment. This limitation can be mitigated by creating a digital twin (DT) as a high-fidelity virtual replica of the physical network environment. By training the policy in the DT, the learning efficiency can be significantly improved~\cite{1_17mihai2022digital}. Recent studies have demonstrated the effectiveness of using DTs for DRL training in wireless networks~\cite{1_18huang2024digital,31li22digital,32guoa23lloca}, revealing that the discrepancy between the DT and reality can be reduced by continuously collecting more real-world data from the actual network. However, the mismatch between the DT and the real environment are practically unavoidable due to incomplete system knowledge, environmental noise delayed transmission, or uncertain network dynamics. These inaccuracies accumulate over time, leading to a growing deviation between the DT and the real environment, which causes LE-UAVs to make suboptimal decisions in trajectory planning and resource allocation. Motivated by the above observations, this paper aims to address the following key challenges:
\begin{itemize}
    \item {Intelligent eavesdropping by the EA-UAV:} The EA-UAV can adaptively adjust its eavesdropping strategy according to LE-UAVs' behavior, making static and pre-defined LE-UAV configurations highly vulnerable. This necessitates more adaptive and resilient operational modes for LE-UAVs to ensure secure data collection from GUs.
    \item {Highly dynamic network environment:} GUs' traffic demands are unknown and highly dynamic, making direct policy deployment ineffective. This calls for a DT model to enable efficient offline policy training and rapid adaptation to real-world dynamics.
    \item {Mismatch in the DT:} Inevitable mismatch between the DT and the real environment degrades the execution performance of the learned policy. A robust mechanism is therefore required to mitigate the impact of such discrepancies during deployment.
\end{itemize}

\subsection{Solutions and Contributions}
In this paper, we focus on a multi-UAV-assisted wireless network operating under the threat of an intelligent EA-UAV.  To enhance both operational flexibility and transmission security, each LE-UAV is capable of dynamically switching between data transmission mode and active jamming mode. This adaptability enables the LE-UAVs to fight against the EA-UAV's eavasdropping more effectively. For instance, when the EA-UAV approaches an data communication link, the actively transiting LE-UAV can switch to active jamming mode, emitting artificial noise to degrade the EA-UAV's channel quality. Conversely, when the EA-UAV moves out of the effective eavasdropping range and the security threat diminishes, the LE-UAV can switch back to data transmission mode to maintain network throughput. Moreover, LE-UAV with sufficient energy resources but low traffic demand can be strategically assigned to aggressively track and continuously jam the EA-UAV. Such a collaborative control among LE-UAVs enables a more efficient trade-off between resistance to eavesdropping attacks and overall network throughput.

To capture the strategic interaction between  LE-UAVs and the EA-UAV, we model this adversarial interaction as a multi-stage Stackelberg game, where the LE-UAVs act as leaders and the EA-UAV serves as the follower.  The secure communication performance is enhanced by jointly optimizing GUs' transmission control, UAV trajectory planning, mode selection, and network formation. This optimization problem involves coupled discrete and continuous decision variables, requires coordinated multi-agent control, and must adapt to the GUs' uncertain traffic demands and the dynamic threat posed by the adversarial EA-UAV.
DRL offers a promising model-free solution by enabling LE-UAVs to learn effective policies through direct interaction with the environmental without prior knowledge. However, conventional DRL suffers from slow convergence due to extensive trial-and-error exploration in the real network. During the learning phase, suboptimal actions may expose communication links to significant security risks, making the real-world action exploration impractical and potentially costly. 

To accelerate training with minimal operational risk, we propose a DT enabled simultaneous learning and modeling (DT-SLAM) framework. In this framework, the DT provides a safe and high-fidelity virtual environment for DRL agent's efficient training. Concurrently, the execution of learned policies in the real network generates valuable feedback data that is used to refine and update the DT modeling, creating a closed-loop co-evolution between DRL learning and DT modeling. Specifically, the DT-SLAM framework relies on LE-UAVs' historical sensing data to construct a dynamic DT model that characterizes the spatial-temporal distribution of GUs' traffic demands and the EA-UAV's eavasdropping behavior. This provide a high-fidelity virtual environment for accelerating policy training. As the LE-UAVs execute policies in the real network, their real-time observations and network feedback are continuously integrated into the DT model, enabling it to adapt and evolve in tandem with the actual network and the agents’ learning progress. By performing fast DRL training in the DT and deploying the converged policy in the real environment, the DT-SLAM framework significantly reduces both learning costs and operational security risks. 

We design a robust proximal policy optimization (RPPO) algorithm for policy learning within the DT. We formulate two asynchronous procedures of learning and modeling operating at different timescales, i.e., one for learning secure transmission strategies within the DT and the other for updating the DT model in response to real network dynamics. 
To enable a lightweight DT, we employ Gaussian process regression (GPR) to predict the key network states required for DRL agent training. The GPR also estimates the model mismatch between the DT and the real environment, which guides LE-UAVs to explore information scarce regions, thereby enabling more accurate DT reconstruction.
Specifically, the main contributions of this paper are summarized as follows:
\begin{itemize}
\item {Game modeling for secure multi-UAV communications}: LE-UAVs can dynamically switch between jamming and data transmission modes in respone to real-time network conditions. The EA-UAV can adapt its eavesdropping strategy to maximize interception performance. We model such dynamic interactions between the LE-UAVs and the EA-UAV as a multi-stage Stackelberg game. We maximize the secure throughput by jointly optimizing GUs' transmission control, UAVs' trajectory planning, mode selection, and network formation strategies.
\item {DT-SLAM for fast learning}: The DT-SLAM framework integrates DRL learning and DT modeling in a dynamically evolving system. It first provides the DT for fast and safe DRL training of the LE-UAVs' secure transmission policies. It also allows continously updates to the DT model by executing the LE-UAVs' actions in reality and simultaneously collecting real-world network feedback.  Their closed-loop integration in DT-SLAM accelerates convergence and enables adaptive, robust decision-making in uncertain and dynamic networks.
\item {Robust learning with DT model mismatch}: We present the RPPO algorithm to improve both the fidelity of the DT and the performance of the learned policies. We employ GPR to estimate the discrepancy between the DT and the real environment and design an uncertainty-aware reward function for PPO agent. This enables the learned policy to better explore the real environment. Experimental results show that RPPO effectively improves the accuracy of the DT and enhances secure transmission performance in the real environment.
\end{itemize}

Preliminary results have partially  been presented in the conference version~\cite{yuan-icc2025}. This paper extends it by investigating secure transmission strategies for LE-UAVs under more sophisticated eavesdropping attacks from the EA-UAV, which can dynamically adjust its trajectory and eavesdropping policy.
We propose a mode-switching scheme that enables each LE-UAV to adaptively alternate between active jamming and data collection to improve overall secrecy performance. Moreover, we develop the DT-SLAM framework to jointly integrate DT modeling and DRL within a dynamically evolving system. This framework is generalizable and can be tailored to various DRL-based control problems in dynamic wireless networks. The rest of this paper is organized as follows. Section~\ref{2} reviews the related works. Section~\ref{chapter3} presents the system model. Section~\ref{4} formulates a multi-stage Stackelberg game formulation. Section~\ref{5} introduces the DT-SLAM framework and details the DRL algorithm for fast and robust  control. Section~\ref{6} presents numerical results, and Section~\ref{7} concludes the paper.

\section{Related Works}\label{2}
\subsection{Secure Communication in UAV-assisted Networks}
UAV-assisted communications are inherently vulnerable to eavesdropping by potential adversaries due to the broadcast nature of wireless signals. A promising countermeasure is physical-layer security. In~\cite{22ding2024collaborative}, the authors addressed active aerial eavesdropping in UAV-enabled mobile edge computing systems by jointly optimizing GUs' communication scheduling, transmission power, and data offloading strategies, as well as the UAV's trajectory planning to ensure secure task execution. In~\cite{10shang2024ris}, the authors employed reconfigurable intelligent surfaces (RIS) to enhance security by adjusting the channel environment. They jointly optimized the  GUs' transmission power, the RIS's phase shifts, and the UAVs' trajectories to strengthen legitimate communication links while suppressing eavesdropper and interferer links. In addition, UAVs can also disrupt eavesdroppers by transmitting jamming signals to increase their noise levels~\cite{12liu2023uav,27wang2024smart,28liao2024game}.
Most existing works assume stationary or predictable eavesdroppers. However, in real-world environments, EA-UAVs can act more intelligently by dynamically adjusting their positions to optimize eavesdropping performance, thereby posing significant security challenges. In~\cite{19wu2022eaves}, the authors showed that EA-UAVs increase their eavesdropping rates by intelligently adjusting trajectories, while LE-UAVs counteract this threat through evasion maneuvers and transmission power adjustment. In~\cite{27wang2024smart}, the authors studied cooperative models where ground jammers and transmission base stations (BS) confront EA-UAVs by generating specific jamming shields between EA-UAVs and legitimate users. They assume independent operation of data transmission and jamming devices, which proves ineffective against adaptive eavesdroppers that can evade jammers and approach transmitters. To enhance the resilience of LE-UAVs against intelligent EA-UAVs, this paper introduces a mode-switching scheme that enables LE-UAVs to dynamically switch between jamming and data transmission modes for efficient secure communications.

\subsection{Deep Reinforcement Learning for Secure Transmissions}
UAV-assisted wireless communication networks are highly dynamic and uncertain, rendering conventional model-based optimization methods ineffective for online and efficient solution~\cite{29lu23RL}. DRL  is well suited to address secure transmission challenges in such networks.
In~\cite{20zhao2024secure}, the authors employed twin delayed deep deterministic policy gradient (TD3) to optimize UAVs' control strategies in UAV-assisted secure video offloading systems under random eavesdroppers. In~\cite{30zhang20madrl}, the multi-agent deep deterministic policy gradient (MADDPG) algorithm was applied to jointly optimize the trajectories and transmission power of jamming UAVs and transmitting UAVs, thereby defending against ground-based eavesdroppers and maximizing secure communication capacity.
Additionally, DRL can effectively address secure communication against intelligent eavesdroppers by dynamically observing their strategies and adapting its own policy accordingly. In~\cite{1_15li2024secure}, the authors addressed EA-UAVs in UAV-assisted mobile edge computing (MEC) using the multi-agent proximal policy optimization (MAPPO), alternately optimizing the policies of legitimate UAVs and the EA-UAV until reaching a dynamic equilibrium. In~\cite{1_16chen2024anti}, the authors modeled an integrated sensing and communication (ISAC) system under malicious jamming as a Stackelberg game. They  employed a game-guided DRL strategy where game theory governs channel selection and DRL optimizes power control to maximize system performance.
Although DRL can effectively solve complex optimization problems, its deployment for training in real-world environments requires extensive interactions with the environment, leading to significant consumption of communication resources, energy, and time. To address this challenge, we propose the DT-SLAM framework to enable low-cost and lightweight DRL training.

\subsection{DT-Enabled Fast Learning for UAV-assisted Networks}
DTs in UAV-assisted networks create virtual replicas of physical systems that enable optimization of network operations and resource management without requiring interaction with the real environment. For example, in UAV-assisted MEC, a DT can be deployed on BS to enable bidirectional data exchange that optimizes task offloading and significantly reduces communication overhead~\cite{31li22digital}. For high-density UAV scenarios, DT edge networks reduce physical interaction requirements while improving quality of service (QoS)~\cite{32guoa23lloca}. The authors in~\cite{34xie2023radar} explored the implementation of DT technology in UAV networks, leveraging RF techniques to process 3D mmWave radar imaging and construct channel models within the DT environment, which supports subsequent resource scheduling tasks. In~\cite{35tang2024uav}, the authors addressed the DT synchronization delay caused by transmitting large volumes of sensor data. They proposed deploying deep neural networks on UAVs and edge servers to extract the semantics of the transmitted information, thereby enhancing the system's real-time performance. The authors in~\cite{36zhou2024cooperative} proposed a double scale spatial DT mapping approach, constructing a large-scale DT to monitor global changes and a small-scale model to provide real-time customized services for UAVs. This method not only reduces the latency of the DT but also enhances its robustness across various scenarios.
Recent studies typically constructed a fixed DT, which leads to a growing mismatch between the digital twin and the real environment as the network dynamics evolve~\cite{1_20shen2023multi}. To address the critical challenge of virtual-physical environment mismatches, the authors in~\cite{33liadaptive} incorporate estimation errors into optimization constraints, enabling DRL to adjust parameters during DT construction. The authors in~\cite{1_22bui2024value} used extended Kalman filtering for DT construction in cloud computing and dynamically update the DT based on estimation errors to better fit the real environment.  To improve DT fidelity, we develop an uncertainty-aware RPPO algorithm that leverages the uncertainty in the DT environment to incentivize the PPO agent to explore the real environment. As such, the PPO agent learns more accurate control policies in the DT that achieve better performance when deployed in the real world.

\begin{figure}[t]
	\centering
	\includegraphics[width=0.98\linewidth]{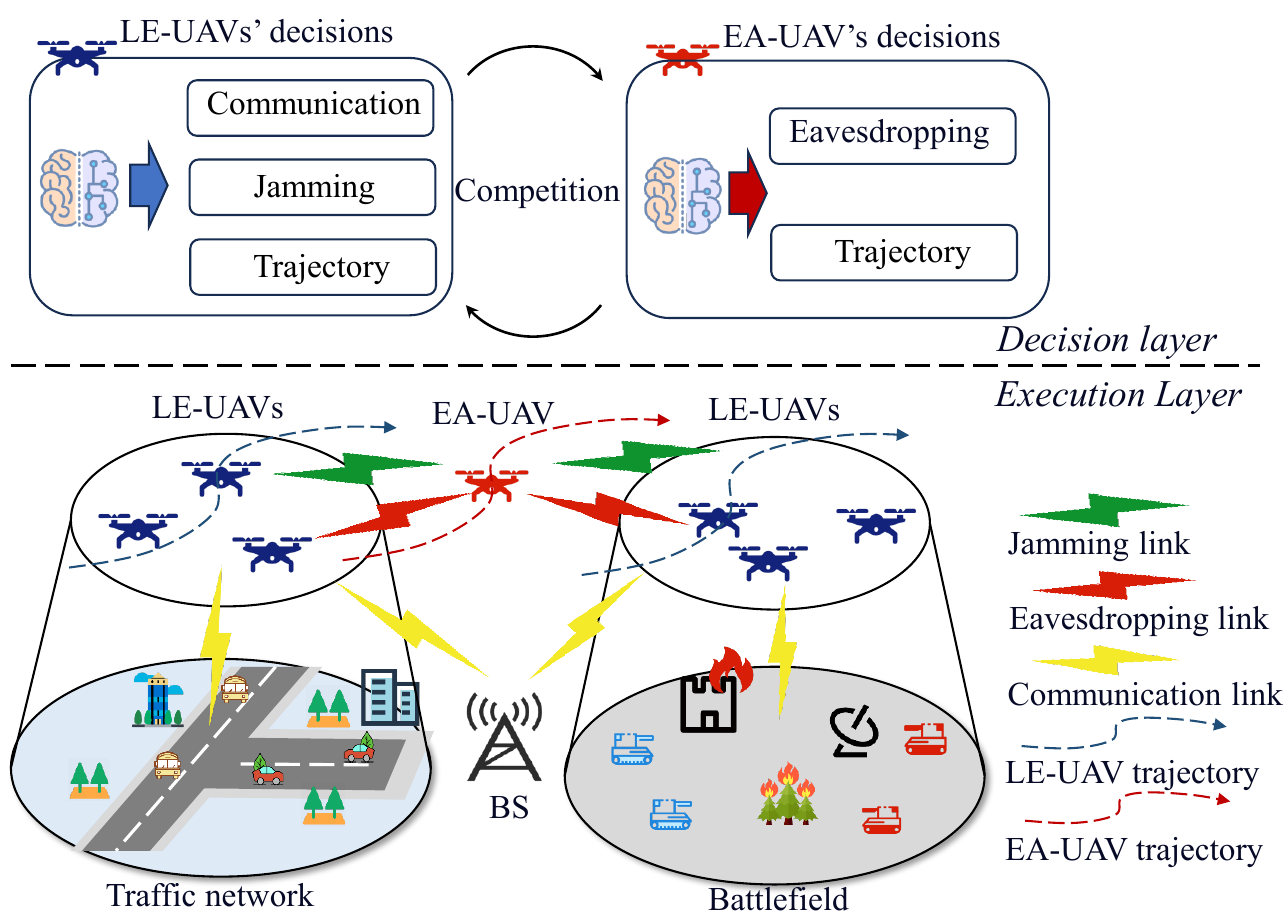}
	\caption{Multi-UAV-assisted secure communications.}
	\label{system_model}
\end{figure}
\section{System Model}\label{chapter3}
As shown in Fig.~\ref{system_model}, we consider a multi-UAV-assisted secure sensing network, including two sets of GUs and LE-UAVs denoted by $\mathcal{Q} = \left\{ 1, 2, \ldots, Q \right\}$ and $\mathcal{Z} = \left\{ 1, 2, \ldots, Z\right\}$, respectively. Both GUs and LE-UAVs are equipped with a single antenna.  All LE-UAVs collaboratively collect data from the GUs and forward it to the BS either directly or via multi-hop relaying through other LE-UAVs. The U2U connections among LE-UAVs can be viewed as the network formation strategy, which can be adaptively reconfigured based on LE-UAVs' trajectories to improve both sensing coverage and transmission efficiency~\cite{6gong2023bayesian}. Besides, an EA-UAV operates within the same service area and attempts to intercept data transmissions from either GUs or LE-UAVs. As such, the LE-UAVs aim to improve secure sensing performance while suppressing the EA-UAV’s eavesdropping capability.

\subsection{Time-slotted Flying, Sensing, and Forwarding}

We adopt a time-slotted frame structure, where the set of time slots is denoted by $\mathcal{T} \triangleq \{1, 2, \cdots, T\}$. Each time slot $t \in \mathcal{T}$ for the LE-UAVs is divided into three sub-slots, i.e., flying sub-slot $t_f^l$, data collection sub-slot $t_c^l$, and data forwarding sub-slot $t_r^l$. During each slot $t$, each LE-UAV first flies to a designated location in $t_f^l$, collects sensing data from GUs in $t_c^l$,  and then forwards the collected data to the BS in $t_r^l$. The UAV–GU associations during the data collection sub-slot $t_c^l$ are represented by a binary matrix $X(t)=[x_{q,z}(t)]_{q \in \mathcal{Q},z \in \mathcal{Z}}$, where $x_{q,z}(t)=1$ indicates that GU-$q$ is scheduled to transmit to LE-UAV-$z$. In the data forwarding sub-slot, each LE-UAV can either transmit its data directly to the BS or relay it through other LE-UAVs.  This collaborative relaying is governed by a dynamic network formation strategy. To unify the notation, we index the BS as LE-UAV-$0$ with the fixed location $\boldsymbol{\ell}_0$. We define the extended node set $\mathcal{F}=\mathcal{Z} \cup \{0\}$, which includes all LE-UAVs and the BS. The network topology during $t_r^l$ is described by a binary adjacency matrix $\Phi(t)=\{\phi_{z,f}(t)\}_{z \in \mathcal{Z},f \in \mathcal{F}}$. For any $z \neq f$, an active communication link exists between LE-UAV-$z$ and LE-UAV-$f$ in the sub-slot $t_r^l$ if $\phi_{z,f}(t)=1$. For simplicity, we assume that  GUs can transmit to at most one LE-UAV and each LE-UAV only forwards its data to at most one next-hop node, which implies the following scheduling and network formation constraints:
\begin{equation}\label{equ-scheduling}
\sum_{z \in \mathcal{Z}}x_{q,z}(t) \leq 1 \text{ and } \sum_{f \in \mathcal{F}}\phi_{z,f}(t) \leq 1.
\end{equation}

The EA-UAV is also assumed to operate in a time-slotted manner. Each operational time slot consists of a flying sub-slot $t_f^e$ and an eavesdropping sub-slot $t_m^e$. The EA-UAV first flies to an eavesdropping location during $t_f^e$ and intercepts potential data transmissions from either GUs or LE-UAVs in the sub-slot $t_m^e$. For simplicity, we assume that the EA-UAV’s time-slotted operation is synchronized with that of the LE-UAVs. Specifically, we set $t_f = t_f^l = t_f^e$, implying that the EA-UAV begins its eavesdropping activity precisely when GUs or LE-UAVs are actively transmitting sensing data.

\subsection{Active Jamming Against Mobile Eavesdropping}

The LE-UAVs' trajectory planning in the first sub-slot $t_f^l$ not only ensures collision avoidance but also aims to maximize the service capacity for all GUs. For notational convenience, we refer to the $z$-th LE-UAV as UAV-$z$ for $z\in\mathcal{Z}$, and denote the EA-UAV as UAV-$e$. For all $i \in \mathcal{Z} \cup \{e\}$, the UAV-$i$' strajectory is defined as a sequence of locations $\mathcal{L}_i=\{\boldsymbol{\ell}_i(1), \ldots,\boldsymbol{\ell}_i(t)\}$ in different time slots. Without loss of generality, we assume that all UAVs maintain a consistent altitude $H$. The GU-$q$'s location is given by $\boldsymbol{k}_q$. For safe operations, all UAVs' trajectories satisfy the following constraints to avoid collisions with the maximum speed limit:
\begin{subequations}\label{equ-cons-v}
\begin{align}
&\| \boldsymbol{\ell}_i(t) - \boldsymbol{\ell}_{i'}(t) \| \geq d_{m}, \quad \forall i,i' \in \mathcal{Z} \cup \{e\} \text{ and } i\neq i', \\
&\| \boldsymbol{\ell}_i(t+1) - \boldsymbol{\ell}_i(t) \| \leq t_{f}\,v_{m}, \quad \forall i\in \mathcal{Z} \cup \{e\},
\end{align}
\end{subequations}
where $d_{m}$ denotes the minimum distance between UAVs, and $v_{m}$ represents UAVs' maximum speed.

The EA-UAV aims to eavesdrop on transmissions from either GUs or LE-UAVs. To enhance secure transmission performance, we allow the LE-UAVs to dynamically switch into an active jamming mode, in which they beamform artificial noise toward the EA-UAV to disrupt its reception. The joint design of active jamming and trajectory planning significantly improves the system’s robustness against eavesdropping attacks.
Let the binary matrix $\Psi(t)=\{\psi_z(t)\}_{z \in \mathcal{Z}}$ denote the operating mode of each LE-UAV operating in  data transmission or jamming modes. When $\psi_z(t) = 1$, UAV-$z$ emits artificial noise to degrade the EA-UAV’s eavesdropping capability. When $\psi_z(t) = 0$, UAV-$z$ participates in legitimate data forwarding. Specifically, the mode selection can be constrained as follows:
\begin{equation}\label{model-switch1}
    \psi_z(t) + \sum_{q \in \mathcal{Q}}x_{q,z}(t) + \sum_{f \in \mathcal{F}}\phi_{z,f}(t) \leq 1, ~\forall z \in \mathcal{Z}.
\end{equation}
By dynamically switching LE-UAVs between jamming and data transmission modes, the system leverages multi-UAV collaborative control to enhance secrecy rate performance, by adapting to instantaneous channel conditions, traffic demands, and resource availability.

\subsection{EA-UAV's Eavesdropping Rate}

\subsubsection{Eavesdropping GUs' transmissions}
Let $d_{q,i}(t)=||\boldsymbol{\ell}_i(t)-\boldsymbol{k}_q||$ denote the distance between UAV-$i$ and GU-$q$ at the $t$-th time slot. The G2U channel between them is modeled as ${h}_{q,i}(t)=\sqrt{\beta_{q,i}} \tilde{h}_{q,i}(t)$, where $\beta_{q,i}=\omega_0{(d_{q,i}(t))}^{-\alpha}$ represents large-scale fading and $\tilde{h}_{q,i}(t) = \sqrt{\frac{K}{1+K}} {h}_{q,i}^{\text{LoS}}+\sqrt{\frac{1}{1+K}}{h}_{q,i}^{\text{NLoS}}$ models small-scale fading with both line-of-sight (LoS) and Non-LoS components. The Rician factor $K$ determines the relative weights of LoS and Non-LoS components. Thus, the transmission rate $R_{q,z}(t)$ from GU-$q$ to UAV-$z$ is modeled as follows:
\begin{equation}\label{equ-uav-rate}
    R_{q,z}(t)=x_{q,z}(t)\text{log}\left(1+\frac{p_{q}|h_{q, i}(t)|^2}{\sigma_q^2 + I_{q,z}(t)  }\right),
\end{equation}
where $\sigma_q^{2}$ is the noise power and $p_{o}$ indicates GU-$q$'s constant transmit power. The term $I_{q,z}(t)$ denotes the interference power received from the other GUs and thus given by:
\begin{equation}\label{equ-gu-intef}
I_{q,z}(t)=\sum_{p \neq q, p \in \mathcal{Q}} x_{p,z}(t)p_{o}|{h}_{p,z}(t)|^2.
\end{equation}
Note that the LE-UAVs' active jamming signal may also introduce inteference power to UAV-$z$'s signal reception. In this paper, we assume that the LE-UAVs' jamming signals are pre-known to all LE-UAVs and thus can be eliminated in~\eqref{equ-uav-rate}. The UAV-$e$'s eavesdropping rate from GU-$q$'s data transmission is represented as follows:
\begin{equation}\label{equ-eav-rate}
    R_{q,e}(t)=x_{q,z}(t)\text{log}\left(1+\frac{p_{q}|h_{q,e}(t)|^2}{\sigma_q^2 + I_{q,e}(t) + G_{q,e}(t)}\right),
\end{equation}
where $I_{q,e}(t)$ is the GUs' inference power, similarly defined as that in~\eqref{equ-gu-intef}, and the extra term $G_{q,e}(t)$ denotes the inteference power induced by the LE-UAVs' active jamming signals, which is represented as follows:
\begin{equation}\label{equ-jamming}
G_{q,e}(t) = \sum_{j \in \mathcal {Z}, j\neq z}\psi_{j}(t) {p}_j |{{h}_{j, e} (t)}|^2,
\end{equation}
where ${p}_j$ is the jamming power. Note that the LE-UAVs aim to collect as much sensing data as possible from the GUs and may approach them closely to improve the transmission rate $R_{q,z}(t)$. However, the EA-UAV may simultaneously move closer to the LE-UAVs, thereby increasing its eavesdropping rate $R_{q,e}(t)$. This motivates the LE-UAVs to dynamically switch to jamming mode to suppress eavesdropping.

\subsubsection{Eavesdropping LE-UAVs' transmissions} In the data forwarding sub-slot $t_r^l$, the data rate from UAV-$z$ to UAV-$f$ is given as follows:
\begin{equation}\label{equ-uav-rate1}
S_{z,f}(t)=\phi_{z,f}(t)\text{log}\left(1+\frac{{p}_{z}{|{h}_{z,f}(t)|^2}}{\sigma_z^{2}+I_{z,f}(t) }\right),
\end{equation}
where ${p}_z$ is the UAV-$z$'s transmit power and $I_{z,f}(t) = \sum_{z^{'} \neq z, z^{'} \in \mathcal{Z}} \phi_{z^{'},f}(t)p_{z^{'}}|{{h}_{z^{'}, f} (t)}|^2$ is channel interference from the other UAVs transmitting simultaneously. Similarly to~\eqref{equ-uav-rate}, we assume that the LE-UAVs' jamming signals are pre-known to UAV-$f$ and thus can be eliminated in~\eqref{equ-uav-rate1}. Similarly, the EA-UAV may eavesdrop the LE-UAVs' forward data transmission. The eavesdropping rate from UAV-$z$'s data transmission can be evaluated as follows:
\begin{equation}\label{equ-eav-rate1}
     S_{z,e}(t)=\phi_{z,f}(t)\text{log}\left(1+\frac{p_{z}{|{h}_{z,e}(t)|^2}}{\sigma_l^{2}+I_{z,e}(t)+G_{z,e}(t)}\right),
\end{equation}
where $I_{z,e}(t)$ is the LE-UAVs' interference at the EA-UAV and $G_{z,e}(t)$ denotes the interference power induced by the LE-UAVs' active jamming signals, similarly to~\eqref{equ-jamming}. The detailed expression is omitted here for brevity.
Therefore, combining~\eqref{equ-eav-rate} and~\eqref{equ-eav-rate1}, the UAV-$e$'s overall eavesdropping rate in the $t$-th time slot can be expressed as follows:
\begin{equation}\label{equ-sum-eave-rate}
\mathcal{E}(t)=\sum_{q\in\mathcal{Q}} t_c^{l}R_{q,e}(t)+\sum_{z\in\mathcal{Z}} t_f^{l}S_{z,e}(t).
\end{equation}
The eavesdropping rate $\mathcal{E}(t)$ measures the data leakage during the data transmissions of both GUs and LE-UAVs.
\subsection{LE-UAVs' Secrecy Rate Peformance}

Let $W_q(t)$ represent GU-$q$'s data queue, which decreases with LE-UAVs' data collection and increases by a random amount $d_q$ per time slot. Thus, the GU-$q$'s data queue evolution is given by:
\begin{equation}\label{equ-gn-buffer}
W_q(t+1)=\left[W_q(t)-\sum_{z \in \mathcal{Z}}t_c^lR_{q,z}(t)\right]^{+}+d_q,
\end{equation}
where $[X]^+=\max\{0, X\}$ represents the max operation. In the $t$-th time slot, the amount of data received by the UAV-$z$ from GUs is $\sum_{q \in \mathcal{Q}}  t_c^{l}R_{q,z}(t)$. Therefore, the UAV-$z$'s data buffer $D_z(t)$ will be evolved as follows:
\begin{align}\label{equ-uav-buffer}
    D_z(t+1)= & \left[D_z(t)- \sum_{f \in \mathcal{Z}\setminus \{ z \}   } t_r^{l}S_{z,f}(t)\right]^+\!   \nonumber \\
    & + \sum_{q \in \mathcal{Q}}  t_c^{l}R_{q,z}(t) +\sum_{z^{'} \in \mathcal{Z}\setminus \{ z \}   } t_r^{l}S_{z^{'},z}(t).
\end{align}
    The data buffer $D_z(t)$ is limited by the LE-UAVs' maximum buffer capacity $D_{x}$. LE-UAVs will discard excessive data if the data size $D_z(t)\ge D_{x}$. The overall throughput received by the BS in the $t$-th time slot is evaluated as follows:
\begin{equation}
u(t) = \sum_{z \in \mathcal{Z}}t_r^{l}S_{z,0}(t).
\end{equation}

Considering the EA-UAV's eavesdropping risk, we aim to maximize the LE-UAVs' overall throughput while minimize the EA-UAV's eavesdropping rate. To this end, we define the system's secure throughput in the $t$-th time slot as follows:
\begin{equation}\label{equ-sec-rate}
\mathcal{U}(t)=u(t)-\lambda \mathcal{E}(t),
\end{equation}
where $\lambda$ is a weighting parameter that balances the LE-UAVs’ transmission capability against their eavesdropping risk. A higher value of $\lambda$ indicates a more sensitive system against eavesdropping attack.

\section{Stackelberg Game for Secure Throughput Maximization}\label{4}

The interactions between LE-UAVs and EA-UAV can be modeled as a multi-stage Stackelberg game. As shown in Fig.~\ref{system_model}, the Stackelberg game formulation involves two agents, typically a leader and a follower, which make sequential best-response decisions based on each other’s strategies and the prevailing dynamic network conditions. In particular, the LE-UAVs act as the leader and first determine a secure transmission strategy to collect sensing data from the GUs while defending against the EA-UAV's eavesdropping attack.
In response, the EA-UAV, as the follower, adjusts its eavesdropping strategy by optimizing its flight trajectory to maximize eavesdropping throughput. In the next step, the LE-UAVs re-optimize their strategies by considering the GUs' current data demands and the EA-UAV's most recent eavesdropping behavior.


\subsection{Rate Maximization for Leader and Follower}
As the leader, given the EA-UAV's eavesdropping strategy, LE-UAVs focus on maximize the overall secure throughput $\mathcal{U}(t)$ by optimizing the GUs' transmission control $\mathbf{X}(t)\triangleq\big\{x_{q,z}(t)\big\}_{z \in \mathcal{Z},q \in \mathcal{Q}}$, and the LE-UAVs' trajectory planning $\mathbf{L}(t)\triangleq\big\{\boldsymbol{\ell}_{z}(t)\big\}_{z \in \mathcal{Z}}$, network formulation $\mathbf{\Phi}(t)\triangleq\big\{\phi_{z,f}(t)\big\}_{z \in \mathcal{Z},f \in \mathcal{F}}$, as well as mode selection $\mathbf{\Psi}(t)\triangleq\{\psi_z(t)\}_{z \in \mathcal{Z}}$, which can be formulated as follows:
\begin{equation}\label{prob-max}
\max_{\mathbf{X}(t),\mathbf{L}(t),\mathbf{\Phi}(t),\mathbf{\Psi}(t)}~  \mathbb{E}[ \mathcal{U}(t) ] \quad  \text{s.t.}
\,\, \eqref{equ-scheduling}- \eqref{equ-sec-rate}.
\end{equation}
The LE-UAVs' sensing throughput $u(t)$ in the $t$-th time slot depends on the transmission rates $(R_{q,z}(t), S_{z,f}(t))$ in both the data collection and data forwarding phases. The GUs' transmission rate $R_{q,z}(t)$ determines how much data each LE-UAV can collect, which in turn dictates the maximum amount $S_{z,0}(t)$ of data each LE-UAV can forward to the BS. To avoid eavesdropping attacks and reduce $\mathcal{E}(t)$, LE-UAVs can also dynamically plan their sensing trajectories to keep away from the EA-UAV or intelligently switch to jamming mode to supress its eavesdropping. Therefore, the LE-UAVs' secrecy rate maximization in~\eqref{prob-max} jointly optimize the GUs' scheduling $\mathbf{X}(t)$, UAVs' trajectory planning $\mathbf{L}(t)$, network formulation $\mathbf{\Phi}(t)$, and mode selection $\mathbf{\Psi}(t)$. Note that problem~\eqref{prob-max} is a nonlinear mixed-integer programming problem, which is difficult to solve efficiently. 

Once the LE-UAVs find their optimal strategies via solving problem~\eqref{prob-max}, the EA-UAV can observe the LE-UAVs' strategy update and promptly adjust its flying trajectory $\ell_{e}(t)$ to enhance the eavesdropping throughput by solving the following rate maximization problem:
\begin{equation}\label{prob-eave-rate}
\max_{\ell_e(t)}~\mathbb{E}[ \mathcal{E}(t) ] \quad \text{s.t.} \,\, \eqref{equ-cons-v}, \eqref{equ-eav-rate}, \text{ and }\eqref{equ-eav-rate1}.
\end{equation}
Here, we assume that the EA-UAV is a powerful eavesdropper capable of simultaneously intercepting all data transmissions from  GUs and LE-UAVs along its trajectory.

In the Stackelberg game model, we employ different utility functions to quantify game players' reward under their respective strategies. For notational simplicity, let $\pi_u$ denote the LE-UAVs' secure transmission strategy and $\pi_e$ represent the EA-UAV's eavesdropping strategy. According to respective maximization problems in~\eqref{prob-max} and~\eqref{prob-eave-rate}, the LE-UAVs' utility function over the mission duration can be defined as $U_{u}(\pi_u|\pi_e) =  \mathbb{E}[\mathcal{U}(t)]$, while the EA-UAV's utility function is given by $U_{e}(\pi_e|\pi_u) = \mathbb{E}[ \mathcal{E}(t)]$. Thus, we can denote the UAV-assisted secure transmission game as $\mathbb{G}=\left<\{\mathcal{Z},\{e\}\},\{\pi_u,\pi_e\},\{U_u,U_e\}\right>$, including two sets of players $\{\mathcal{Z},\{e\}\}$ with individuals' control strategies $\{\pi_u,\pi_e\}$ and utility functions $\{U_u,U_e\}$. Both players (i.e., LE-UAVs and the EA-UAV) individually update their strategies to maximize their own utilities in an alternative manner. In particular, given the opponent's strategy, each player chooses its best response to improve the utility by solving the rate maximization problem in~\eqref{prob-max} or~\eqref{prob-eave-rate}.

\subsection{General DRL Approach for Rate Maximization}

Model-free DRL is well suited to complex network environments because it does not require an environment model or prior knowledge. It learns optimal strategies through trial and error by using only observed states. Thus, we first reformulate problems \eqref{prob-max} and \eqref{prob-eave-rate} into markov decision process (MDP), providing a mathematical modeling for sequential decision-making. MDP is defined by the triples $<\mathcal{S},\mathcal{A},\mathcal{R}>$, representing state, action, and reward spaces, respectively.

For the LE-UAVs' secrecy rate maximization in~\eqref{prob-max}, the state in the $t$-th time slot consists of all UAVs' information ${\mathbf s}^l(t)=\big\{ \mathbf{L}(t),\mathbf{R}(t),\mathbf{S}(t),\mathbf{D}(t) \big\}$, including all UAVs' locations
$\mathbf{L}(t)=\{\boldsymbol{\ell}_i(t)\}_{i \in \mathcal{I}}$, LE-UAVs' data buffer $\mathbf{D}(t)=\{D_z(t)\}_{z \in \mathcal{Z}}$, as well as the UAVs' transmitting rates $\mathbf{R}(t)=\{R_{q,z}(t)\}_{q \in \mathcal{Q},z \in \mathcal{Z}}$ and $\mathbf{S}(t)=\{S_{z,f}(t)\}_{z \in \mathcal{Z},f \in \mathcal{F}}$ in the data sensing and forwarding phases. The LE-UAVs' action $\mathbf{a}^l(t)$ includes the trajectory planning of all LE-UAVs $\mathbf{L}(t)$, the UAV-GU scheduling policy $\mathbf{X}(t)$, the network formation $\mathbf{\Phi}(t)$, and the mode selection $\mathbf{\Psi}(t)$. The reward depends on the state and the LE-UAVs' actions, which is related to the secure throughput reward $\mathcal{U}(t) $ in the optimization objective and defined as follows:
\begin{equation}\label{reward1}
r_c(t)=\mathcal{U}(t)-\mu_1r_f(t)-\mu_2r_v(t),
\end{equation}
where $\mu_1$ and $\mu_2$ are weighting coefficients combining two penalty terms in the reward. The second term   $r_{f}(t)=\sum_{z,z^{'} \in \mathcal{Z}} \mathbf{I}\left(\left\|\boldsymbol{\ell}_z(t)-\boldsymbol{\ell}_{z^{'}}(t)\right\|\leq d_{\min }\right)$ enforces minimum safe distances between UAVs, where $\mathbf{I}(\cdot)$ is an indicator function returning $1$ when constraints are met and $0$ otherwise. The penalty term applies a negative value when UAVs breach the safety threshold. The third term $r_v(t)=\sum_{z \in \mathcal{Z}} \mathbf{I} \left(\left\|\boldsymbol{\ell}_z(t+1)-\boldsymbol{\ell}_{z}(t)\right\|\geq t_fv_m\right)$ limits the LE-UAVs' speed less than the maximum speed $v_m$. 

For the EA-UAV's rate maximization problem in~\eqref{prob-eave-rate}, the state in $t$-th time slot ${\mathbf s}^e(t)$ includes all UAVs' locations $\mathbf{L}(t)$, the eavesdropping rate from GUs to EA-UAV $\mathbf{R}_e(t)=\{R_{q,e}\}_{q \in \mathcal{Q}}$, the eavesdropping rate from LE-UAVs to EA-UAV $\mathbf{S}_e(t)=\{S_{z,e}\}_{z \in \mathcal{Z}}$. The action $\mathbf{a}^e(t)$  consists of EA-UAV's trajectory planning $\ell_e(t)$. The EA-UAV's reward can be expressed as follows:
\begin{equation}
    r_e(t)=\mathcal{E}(t)-\mu_e r_e(t),
\end{equation}
where $\mu_e$ is the weighting coefficient combining the eavesdropping rate $\mathcal{E}(t)$ and the penalty term to ensure operational safety as follows:
\begin{align*}
r_e(t)\,=\, & \mu_1 \sum_{z\in \mathcal{Z}} \mathbf{I}\left(\left\|\boldsymbol{\ell}_e(t)-\boldsymbol{\ell}_{z}(t)\right\|\leq d_{\min }\right)+ \\
& \mu_2 \sum_{z\in \mathcal{Z}} \mathbf{I} \left(\left\|\boldsymbol{\ell}_e(t+1)-\boldsymbol{\ell}_{e}(t)\right\|\geq t_fv_m\right).
\end{align*}

Once formulated into MDP problems, the LE-UAVs' and EA-UAV's rate maximization problems can be solved through interaction with environment using DRL method. For instance, we can directly employ PPO algorithm due to its training stability, which is achieved by clipping the probability ratio to constrain policy updates~\cite{21schulman2017proximal}.

\section{DT-SLAM Framework for Fast Learning under Model Mismatch}\label{5}

Although problems \eqref{prob-max} and \eqref{prob-eave-rate} are fully formulated and can in principle be solved by the DRL, directly training the UAVs in the real environment raises the following concerns:
\begin{itemize}
    \item {High interaction overhead:} DRL requires frequent interactions with the real environment to collect feedback observations and update the policy. In UAV-assisted communication networks, such interactions involve substantial information exchange, consuming considerable spectrum and energy resources. Since both resources are inherently limited in UAV-assisted networks, this requirement poses a significant practical challenge, especially when the network scale is large.
    \item {Inefficient exploration in unknown environments:} The DRL relies on continuous trial-and-error exploration to discover effective strategies, which becomes particularly challenging when the environment dynamics are unknown or highly dynamic. In our scenario, the unpredictable behavior of the EA-UAV and the time-varying traffic demands of GUs make it difficult for LE-UAVs to efficiently explore the state-action space, resulting in an unstable training performance in practice.
\end{itemize}

To develop a practically DRL algorithm that can be efficiently trained and reliably executed in real-world environments, we propose a DT-SLAM framework to address the above limitations. By allowing DRL agents to train within the DT instead of the real environment, the proposed framework significantly reduces real-world interactions, accelerates policy learning, and ensures safe exploration without disrupting normal network operations. To mitigate the potential mismatch between the DT and the real environment, we further introduce a RPPO algorithm, which not only boosts learning efficiency but also actively aligns the DT with the real environment, thereby ensuring robust and adaptive policy learning under practical constraints.


\begin{figure}
    \centering
    \includegraphics[width=0.98\linewidth]{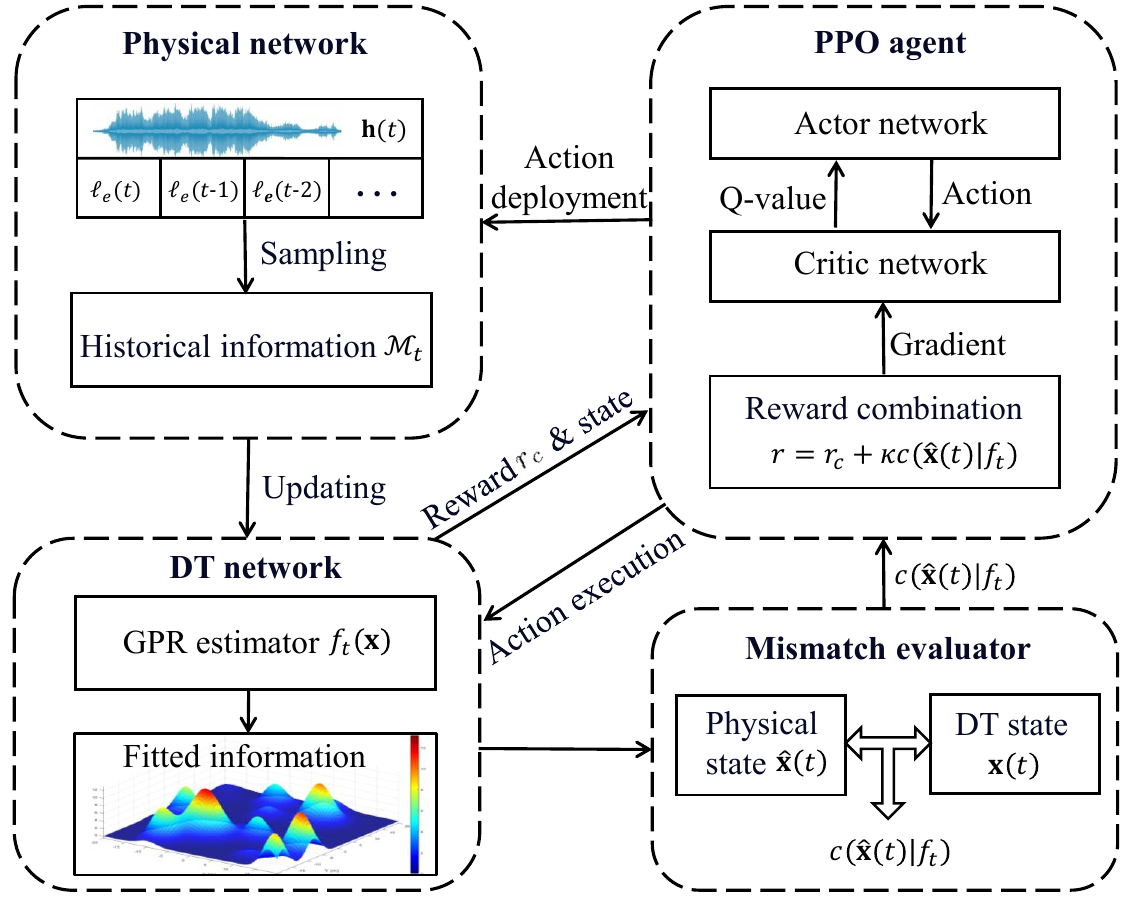}
    \caption{The DT-SLAM framework with RPPO algorithm.}
    \label{fig:DT-RPPO}
\end{figure}

DT technology provides a lightweight virtual replica of the real-world environment. This virtual surrogate enables simulated interactions that replace costly real-world trials,  accelerating DRL convergence.  As shown in Fig.~\ref{fig:DT-RPPO}, the DT-SLAM framework consists of inter-DT learning and intra-DT update loops. A DRL agent is deployed in the inter-DT learning loop to optimize its policy without direct interaction with the real environment. The intra-DT loop updates the DT model based on observations periodically sampled by LE-UAVs.  Specifically, given an up-to-date DT model of the real network, we reformulate the UAVs' control problems as a DT-enhanced MDP. The DRL agent trains within this virtual environment, where rich and low-cost observations facilitate rapid policy learning. Once converged, the learned policy is deployed in the real network, significantly reducing the time and resources otherwise spent on real-environment interactions. Furthermore, the UAVs' execution of actions in the real network environment generates new sensing data, capturing dynamic user traffic demands, time-varying channel states, and even adversarial behaviors such as EA-UAV's eavesdropping. This real-world data is then fed back to refine and update the DT model, ensuring its timely synchronization  aligned with the true dynamics of the network environment.

\subsubsection{Inter-DT fast learning} Starting from an initial DT, the UAVs' control problems are formulated as MDP and solved using DRL algorithms within the virtual environment, enabling rapid convergence without real-world deployment costs. In the inter-DT learning loop, the DRL agent continuously interacts with the DT to optimize secure transmission throughput and explore high-value sensing regions. By accurately modeling and tracking system dynamics, the DT emulates key aspects of the physical network, including time-varying channel and energy states, buffer statuses of both GUs and UAVs, the evolution of GUs' traffic demands, and UAVs' mobility patterns. This allows the DT to predict future network state transitions as the DRL agent executes actions in the virtual environment. Such inter-DT learning loop can be very fast without actual time overhead in obvserving the actual system transitions and the inefficient resource consumption in trials and errors.

\subsubsection{Intra-DT modeling update} An accurate DT is essential for the inter-DT DRL agent to learn policies that generalize well to the real-world network. However, in practice, it is often infeasible or prohibitively costly to obtain complete and precise network information to construct the DT model, particularly in dynamic wireless environments characterized by stochastic channel fluctuations and time-varying traffic demands. These inherent uncertainties necessitate an evolving DT model that can be continuously refined using real-time observations and data samples collected from the real-world network. This requirement motivates the intra-DT modeling update loop, as depicted in Fig.~\ref{fig:DT-RPPO}. Following inter-DT DRL training, the converged policy is deployed in the real network to accumulate secure throughput performance. During this execution phase, the UAVs actively explore the environment and simultaneously gather new data samples, such as channel measurements, traffic patterns, and mobility traces, which are then fed back into the intra-DT learning loop.

Given these newly acquired real-world samples, the DT model can be enhanced through multiple approaches, including data-driven probabilistic estimation or model-free learning techniques. Specifically, a model-free intra-DT learning method can leverage DNNs~\cite{Tang-iotj2024} to approximate the underlying state-transition dynamics. The data collected during DRL policy execution provides additional labeled transition samples (state–action–next-state tuples), thereby improving the fidelity of the learned dynamics model for subsequent inter-DT training iterations. Alternatively, a more lightweight approach employs Bayesian inference for probabilistic modeling. For instance, to estimate GUs' spatial traffic demand distribution, we can maintain a Bayesian posterior over possible demand patterns based on historical sensing data. As LE-UAVs collect fresh traffic observations from GUs, this posterior is updated in real time, yielding an increasingly accurate representation of the GUs' traffic demand distribution, which is sufficient for policy learning in the DT environment.

\subsubsection{Capturing the model mismatch}

The inter-DT DRL agent aims to maximize long-term secure throughput through strategic action selection and environmental exploration. The execution of action in real-world network  also generates valuable data that enhances the DT's accuracy via the intra-DT update loop. This creates a tight coupling between policy learning and model refinement, i.e., the DRL agent not only seeks optimal control decisions but also plays an active role in refining the DT by collecting informative samples during execution. To explicitly account for this tradeoff, we propose quantifying the DT model mismatch, i.e., the discrepancy between the DT's predicted dynamics and actual system behavior, and incorporating this metric as a penalty term in the DRL agent's reward function. As such, the agent is incentivized not only to achieve high throughput performance but also to reduce model mismatch through action exploration, thereby accelerating the co-evolution of the DRL agent's policy and the DT model.


\subsection{Intra-DT Modeling Updates}
In this part, we propose a light-weight implementation for updating the DT model using newly collected sampling data from real-world network interactions. Rather than striving to maintain a fully accurate DT that replicates the entire physical environment, our approach focuses exclusively on capturing the time-varying dynamics of key network factors that directly influence the sensing and transmission performance of LE-UAVs. These factors include GUs' channel conditions ${\bf h}(t)=\{{h}_{q,z}(t)\}_{q\in\mathcal{Q},z\in\mathcal{Z}}$ and the trajectory points of the EA-UAV $\boldsymbol{\ell}_e(t)$. Hence, our light-weight DT leverages data-driven predictive and probabilistic surrogate models to capture the variations of these network factors.
GPR is a non-parametric Bayesian method used for accurate predictions and  uncertainty quantification in data-driven modeling. It treats the underlying dynamics (such as the GUs' channel conditions and UAVs' locations) as a stochastic process, estimating a full posterior distribution over possible functions consistent with observed data.  Since GPR requires no pre-defined model structure, it can be seamlessly applied to the time-varying multi-UAV network without prior knowledge of the environment dynamics.  Furthermore, the uncertainty estimates provided by GPR facilitate our RPPO design to promote more accurate DT modeling.

As illustrated  in Fig.~\ref{fig:DT-RPPO}, GPR is employed at the BS to reconstruct a complete representation of the environmental state from the partial observations reported by LE-UAVs.
Based on current network observations, each LE-UAV leverages prior knowledge to refine its estimates of GUs' channel conditions and the EA-UAV's location. These updated network states are then integrated into the DT and shared among LE-UAVs, serving as informative inputs for inter-DT collaborative learning.
Let ${\boldsymbol \eta}(t)=\{{\bf h}(t), \boldsymbol{\ell}_e(t)\}$ denote the LE-UAVs' interactive observations with the network environment in the $t$-th time slot. Each LE-UAV periodically reports $\boldsymbol{\eta}(t)$ to the BS, enabling the construction of a probabilistic model that captures the underlying dynamics of $\boldsymbol{\eta}(t)$. By the $t$-th time slot, the cumulative historical information is denoted as $\mathcal{M}_t=\{{\boldsymbol\eta(t)}\}_{t \in \mathcal{H}_t}$, where $\mathcal{H}_t$ denotes the finite time window that retains most recent interactions between the LE-UAVs and the real environment. This memory buffer $\mathcal{M}_t$ captures the temporal evolution of the recent network environment.
Leveraging $\mathcal{M}_t$, the DT can update the network state across all locations within the service area. This data-driven  estimation allows us to understand how the network evolves in the near future. Thus, it achieves a light-weight DT environment to accurately reflect anticipated changes of the network states that are closely relevant to the performance optimization problem and the inter-DT DRL's policy learning, without requiring full-scale simulation or exhaustive sensing.

\subsubsection{Preparing data samples during action execution}
The historical data $\mathcal{M}_t$ can be updated each time when the DRL agent executes its action in the real network environment and thus obtain new data samples for updating the DT model. In particular, during action exectuion, each LE-UAV can move to the next trajectory point and receive new sensing data from the GUs. The size of sesning data received at each location can help estimate the geographical distribution of  GUs' traffic demands. Considering the dynamic nature of the UAV-assisted wireless environment, each LE-UAV may revisit the same location or serve the same set of GUs multiple times. Different LE-UAVs may also vist the same location in different time slots on their trajectories. As the network changes rapidly, the LE-UAVs' observations at the same location may be very different to each other. This requires us to design a data fusion mechanism to avoid contradictory data samples and prepare the sampling data for updating the DT model via GPR method.

A simple implementation of the data fusion mechanism is to ensure consistence of the data samples collected in different time slots, i.e., we can simply append the new data samples into the historical dataset $\mathcal{M}_t$ if the new data samples is absent in the previous dataset $\mathcal{M}_{t-1}$, otherwise we replace the obsolete observations by the new sampling data. Specifically, for any $z\in\mathcal{Z}$, let $\delta_{z}(\mathcal{M}_{t-1})$ denote the set of data samples in $\mathcal{M}_{t-1}$ that are collected within a short distance $\delta_{\min}$ with respect to the LE-UAV-$z$'s current location $\boldsymbol{\ell}_z(t)$, as follows:
\begin{equation}
\delta_{z}(\mathcal{M}_{t-1}) \triangleq \left\{ {\bf s}_j \in \mathcal{M}_{t-1} \mid \| \boldsymbol{\ell}_z(t) - \boldsymbol{\ell}_{z,j} \| \leq \delta_{\min} \right\},
\end{equation}
where ${\bf s}_j$ denotes the $j$-th data sample in $\mathcal{M}_{t-1}$ and $\boldsymbol{\ell}_{z,j}$ denotes the location of LE-UAV-$z$ when ${\bf s}_j$ was collected. When two LE-UAVs are close to each other, we assume they can have the same observations about the network conditions. When $\delta_{z}(\mathcal{M}_{t-1})= \emptyset$, it indicates that the LE-UAV-$z$'s new sampling data is collected at a  different location. Hence, we can simply append the new observation $\boldsymbol{\eta}(t)$ into the historical data, i.e.,~$\mathcal{M}_{t} = \mathcal{M}_{t-1} \cup \{\boldsymbol{\eta}(t)\}$. Otherwise, we replace the obsolete observations in $\delta_{z}(\mathcal{M}_{t-1})$ by the new observation $\boldsymbol{\eta}(t)$. As such, for any LE-UAV-$z$'s interactions with the real network, we can formulate the heuristic data fusion in a compact form as follows:
\begin{equation}\label{equ-update}
    \mathcal{M}_t = \mathcal{M}_{t-1} \setminus \delta_{z}(\mathcal{M}_{t-1}) \cup \{\boldsymbol{\eta}(t)\}, \quad \forall z\in\mathcal{Z}.
\end{equation}
The fusion mechanism~\eqref{equ-update}  denotes that the historical observations $\delta_{z}(\mathcal{M}_{t-1})$ in $\mathcal{M}_t$ that are out of date and may be contradictory with the current observation $\boldsymbol{\eta}(t)$. 
Thus, the obsoleted information $\delta_{z}(\mathcal{M}_{t-1})$ should be removed from the historical dataset $\mathcal{M}_t$ to ensure data consistency.

\subsubsection{Updating DT model by GPR}
Each network state is updated in parallel within the DT model.
Let ${\bf x}(t) \in \boldsymbol{\eta}(t)$ denote a latest observation of a specific network variable (e.g., ${\bf h}(t)$ and $ \boldsymbol{\ell}_e(t)$ ) from the real environment.
The network state $\widehat{{\bf x}}(t)$ in the DT is predicted as follows:
\begin{equation}
    \widehat{{\bf x}}(t)= f_t({\bf x}(t))+\boldsymbol{\epsilon},
    \label{function}
\end{equation}
where $\boldsymbol{\epsilon} \sim \mathcal{N}(\mathbf{0}, \sigma_1^2 \mathbf{I})$ denotes independent Gaussian noise with variance $ \sigma_1^2$. The function $f_t(\cdot)$ represents a GPR-based transition function learned from the historical observations $\mathcal{M}_t$, which predicts the GUs' channel conditions and the EA-UAV's position at the next time slot.  Specifically, the function $f_t(\cdot)$ is modeled as a Gaussian process as follows:
\begin{equation}
f_t({\bf x}) \sim \mathcal{N}(m({\bf x}),k({\bf x},{\bf x}')),
\end{equation}
where $m({\bf x})$ is the mean function and $k({\bf x},{\bf x}')$ is the kernel function  that quantifies the correlation between two network states. The inputs ${\bf x}$ and ${\bf x}'$ in the kernel function represent distinct state samples drawn from the historical observation set $\mathcal{M}_t$. Following common practice in Gaussian process modeling, we adopt a zero-mean prior, i.e., $m(\mathbf{x})=0$, similar to that in~\cite{20cheng2022rethinking}. The kernel function employs the squared exponential kernel, i.e., $k({\bf x},{\bf x}')=\alpha^2_t \exp (-\frac{{||{\bf x}-{\bf x}'||}^2}{2l_t^2})$, where $\alpha_t$ controls the overall variance of the predicted network state and $l_t$ is the length-scale parameter that determines how sensitive the learned function is to changes in the input states. Therefore, $\alpha_t$ and $l_t$ shape the smoothness and flexibility of the transition function $f_t(\cdot)$. To ensure $f_t(\cdot)$ accurately reflects the observed dynamics in $\mathcal{M}_t$, we optimize the hyperparameters $\alpha_t$ and $l_t$ by maximizing the marginal likelihood of the historical data as follows:
\begin{equation}
    \alpha_t,l_t = \arg\max_{\alpha_t,l_t}\log p(\mathbf{Y}|\mathbf{X},\alpha,l),
\end{equation}
where $\mathbf{X}$ and $\mathbf{Y}$ denote the matrices  of historical input states and the corresponding output labels both extracted  from the observation set $\mathcal{M}_t$. The function $p(\cdot)$ is the marginal likelihood derived from the Gaussian process prior.
Hence, the DT predicts the network state $\widehat{{\bf x}}(t)$ as follows:
\begin{equation}\label{predict}
    \widehat{{\bf x}}(t) = k({\bf x}(t), \mathbf{X}) \big( k(\mathbf{X}, \mathbf{X}) + \sigma_2^2 \mathbf{I} \big)^{-1} \mathbf{Y},
\end{equation}
where  $\sigma_2^2$ represents the variance of the observation noise.  The detailed derivation follows the approach in~\cite{20cheng2022rethinking} and is omitted here for brevity.
\subsection{Inter-DT Fast Learning under Model Mismatch}
In the DT-SLAM framework, UAVs are trained  within the DT model. We expect that the UAV's policies learned in the DT can be directly deployed in the real environment. However, if the DT differs significantly from the actual network dynamics, the resulting policies may exhibit degraded performance when applied in practice. Therefore, it is important that the DRL's action  applied in DT not only optimizes the target throughput performance but also actively promotes fidelity between the DT and the real-world environment by ensuring accurate modeling of time-varying network dynamics.

Beyond state prediction, GPR also inherently provides uncertainty quantification in the form of predictive variance. This uncertainty estimate reflects the confidence in the DT's reconstruction of the environment, particularly in regions sparsely observed by LE-UAVs. Thus, by incorporating this uncertainty into the DRL reward function, the agents can improve their sampling efficiencies and accelerate adaptations upon real-world deployment.
The uncertainty of network state $\widehat{{\bf x}}(t)$ in DT and real network state in real environment ${\bf x}(t)$ can be calculated via GPR as follows:
\begin{equation}\label{var}
\begin{split}
c(\widehat{{\bf x}}(t)|f_t)&=
k({\bf x}(t), {\bf x}(t)) \\
&- k({\bf x}(t), \mathbf{X}) \big( k(\mathbf{X}, \mathbf{X}) + \sigma_2^2 \mathbf{I} \big)^{-1} k(\mathbf{X}, {\bf x}(t)),
\end{split}
\end{equation}
where $c(\widehat{{\bf x}}(t)|f_t)$ denotes the predictive variance of the DT at the estimated network state $\widehat{{\bf x}}(t)$. A larger value of $c(\widehat{{\bf x}}(t)|f_t)$ indicates a greater mismatch between the predicted DT state and the true state in the real environment.  Intuitively, a high predictive variance signifies that the DT has limited or sparse observational data in the corresponding region of the state space. This suggests that LE-UAVs should prioritize collecting additional real-world samples from these uncertain regions to refine the DT and better capture the underlying network dynamics. To leverage this insight, we propose the RPPO algorithm that explicitly incorporates the predictive uncertainty $c(\widehat{{\bf x}}(t)|f_t)$  into the design of the DRL reward function. The RPPO serves two purposes. First, it guides LE-UAVs to learn efficient and secure data collection strategies within the DT. Second, it encourages LE-UAVs to explore uncertain regions during real-world deployment, facilitating more accurate DT modeling.

Specifically, we adopt PPO as the DRL algorithm to learn the UAVs' control policies within the DT.
The interaction between the UAVs and the DT is modeled as a MDP, defined by the tuple $\langle\widehat{\mathcal{O}}, \widehat{\mathcal{A}}, \widehat{\mathcal{R}} \rangle$, where $\widehat{\mathcal{O}}$ denotes the observation space consisting of DT-predicted network states obtained via GPR.  $\widehat{\mathcal{A}}$ represents the joint action space of the LE-UAVs and $\widehat{\mathcal{R}}$ is the reward function guiding policy learning. For brevity, we omit the time index in the DT formulation in the following. The action $a \in \widehat{\mathcal{A}}$ is defined as $\{\psi_z, \boldsymbol{\ell}_z, x_{q,z}, \phi_z\}_{z \in \mathcal{Z}}$. The reward function $r \in \widehat{\mathcal{R}}$ is defined as follows:
\begin{equation}\label{reward1}
    r=r_c+\kappa c(\widehat{{\bf x}}(t)|f_t),
\end{equation}
where $r_c$ is defined in~\eqref{reward1} to capture the secure throughput performance and the non-negative weight $\kappa \geq 0$ balances the trade-off between task performance and exploratory behavior. Reward~\eqref{reward1} indicates that the LE-UAVs not only aim to improve transmission efficiency within the DT but also are encouraged to explore regions where the DT model exhibits high uncertainty. When deployed in the real-world environment, this exploration drives LE-UAVs to collect informative observations from under-explored regions, which  enables more accurate DT modeling in the subsequent update round.

The DT-SLAM framework with RPPO (denoted by DT-RPPO) is summarized in Algorithm~\ref{alg-gprrppo}, which comprises two nested loops. In the outer loop, the GPR model parameters are initialized and periodically updated using newly acquired real-world channel observations, ensuring aligned DT with the real environment. Within the DT, the inner loop executes the proposed RPPO to optimize the UAVs' control policies. Periodically, the updated policy is deployed on the LE-UAVs, whose interactions with the environment generate fresh data that further refine the DT.  This iterative process continues until ultimately yielding  stable policies for all LE-UAVs operating in the real environment.

\begin{algorithm}[t]
	\caption{DT-RPPO for UAVs' transmission controls}\label{alg-gprrppo}
	\begin{algorithmic}[1]
        \State Initialize the parameters of GPR and DNN for all UAVs. 
        \For{ $t=1,2,\cdots,T$}
        \Statex \textbf{\% Executing  UAVs' policies in real environment}
        \State UAVs perform actions adapted by the PPO
        \State Record the collected data $\delta_2$
        \State Return historical dataset $\mathcal{M}_t$
        \Statex \textbf{\% Synchronize the DT with real environment}
        \State Train the environment GPR estimator $f_t({\bf x})$
        \State Evaluate the model mismatch  $c(\widehat{{\bf x}}(t)|f_t)$
        \State  Update DT via GPR
        \Statex \textbf{\% Optimizing  UAVs' policies in DT}
        \State Obtain the current status $\widehat{{\bf x}}(t)$ of the DT
        \State Execute the action $a$ according to the PPO
        \State Transition to the next state and returns the reward $r$
        \State Update the parameters of actor- and critic-networks
        \EndFor
	\end{algorithmic}
\end{algorithm}

\section{Numerical Results}\label{6}
\begin{table}
\caption{Parameter settings in the simulations.} \label{para_settings}\normalsize
	\centering
	\begin{tabular}{|l|l|}
		\hline
        Parameters&Settings\\
        \hline
		GU's transmit power $p_q$ & $26$ dBm \\
		Flying altitude of the UAVs $H$ & $100$ m\\
        Background noise power $\sigma^2$& $-90$ dBm\\
		Maximum speed of the UAVs $v_{{m}}$& $20$ m/s\\
        Safety distance to the UAVs $d_{{m}}$ & $5$ m\\
        Actor's learning rate& $10^{-4}$\\
        Critic's learning rate&$10^{-3}$\\
        Reward discount factor &$0.95$\\
        Size of the replay buffer &$1500$\\
        Size of the mini-batch &$150$\\
        $\epsilon$-greedy coefficient & $0.1$\\
        \hline
	\end{tabular}
\vspace{-0.6cm}
\end{table}

This section presents numerical results to evaluate the performance of the proposed  RPPO algorithm and the UAV mode-switching jamming strategy in multi-UAV secure communication networks. We consider a scenario with
$M = 3$ UAVs collecting data from $K = 30$ GUs randomly distributed within a $(2000 \times 2000) m^2$ area. The UAV-GU links follow a Rician fading model, , while UAV–UAV communications are assumed to be dominated by LoS propagation. Other simulation parameters are summarized in Table~\ref{para_settings}.

\subsection{Convergence Performance of the RPPO}

\begin{figure}[t]
\centering
    \subfloat[ Secure throughput with different interaction frequency.]{\includegraphics[width=0.98\linewidth]{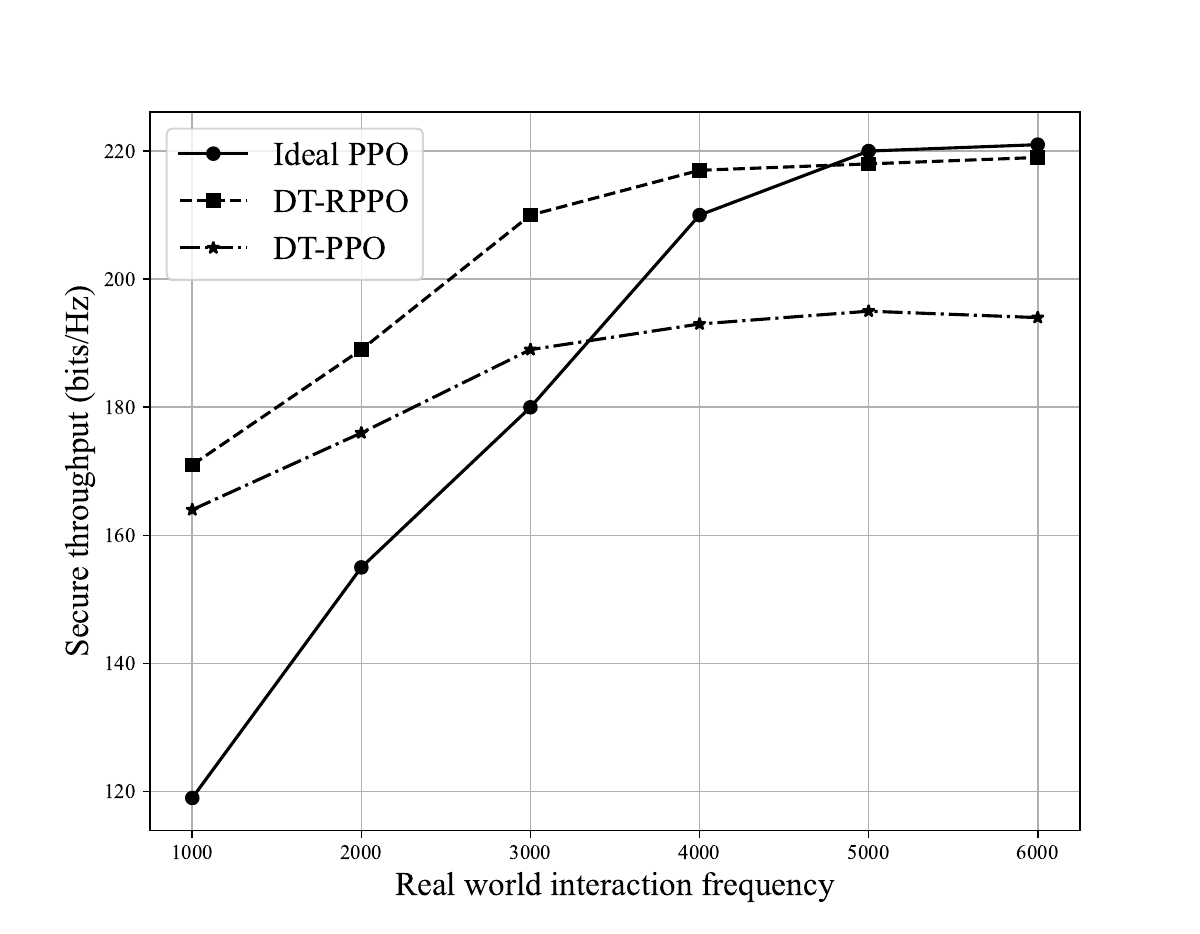}}\\
     \subfloat[Convergence performance of the RPPO algorithm.]{\includegraphics[width=0.85\linewidth]{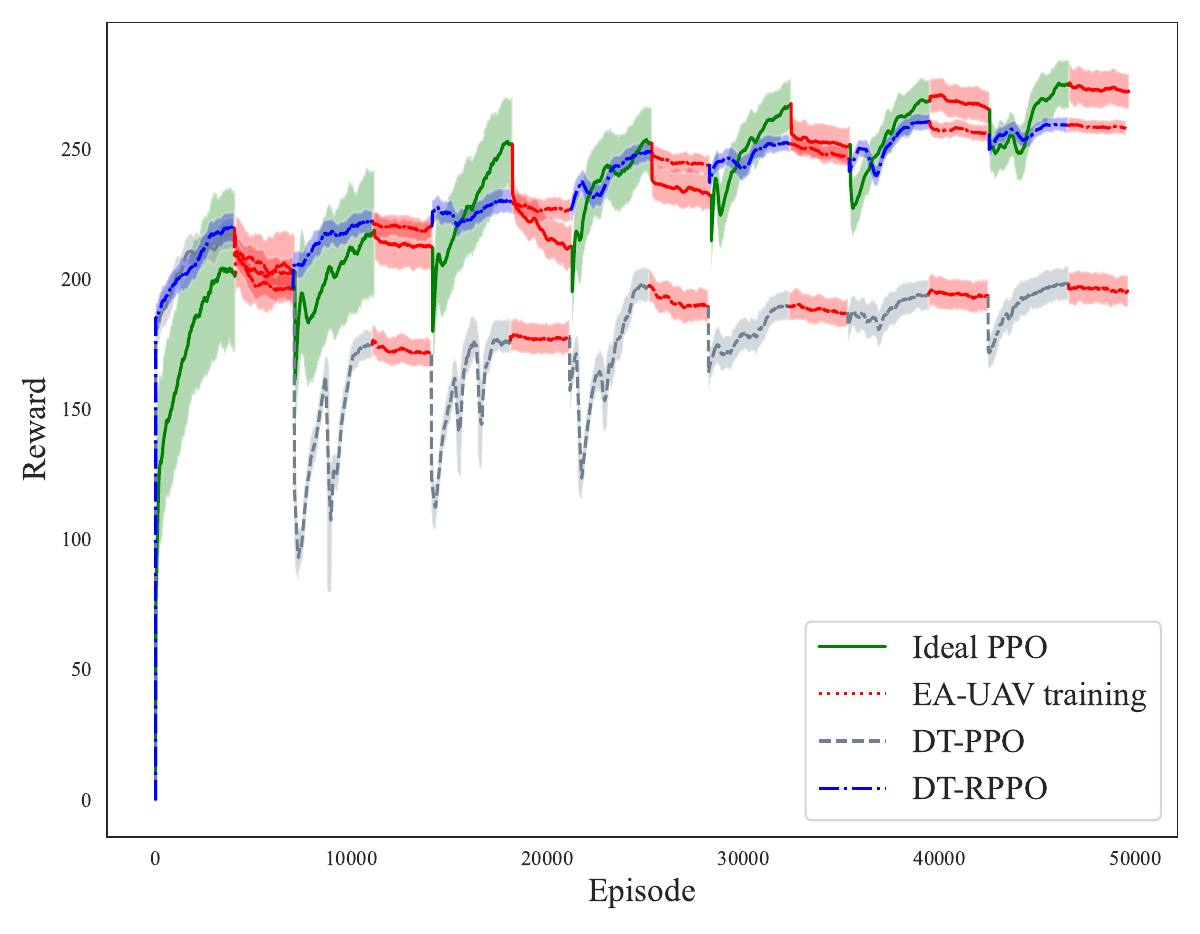}}\\
\caption{DT improves the learning performance.}
\label{dt33}
\end{figure}

\begin{figure}
    \centering
    \includegraphics[width=0.98\linewidth]{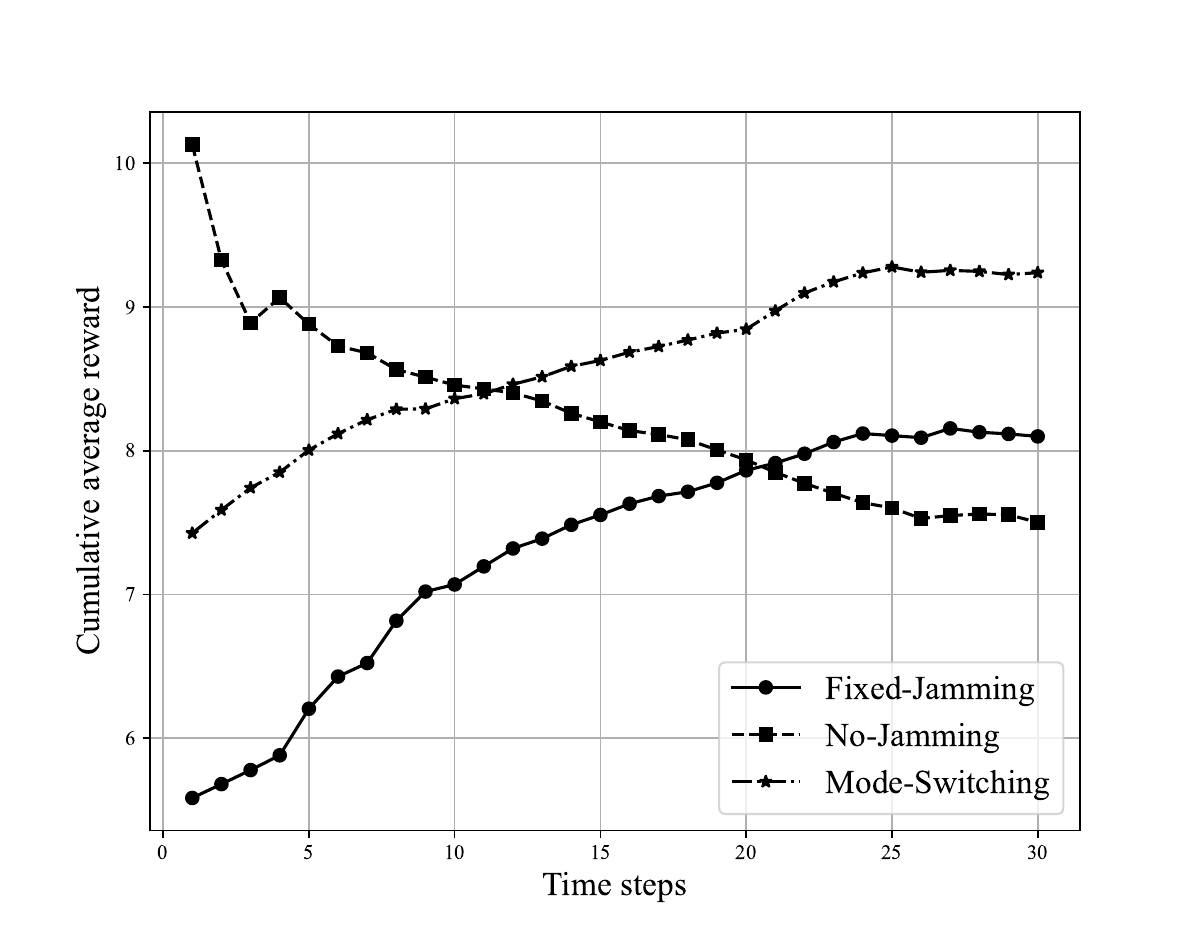}
    \caption{Reward performance in different time slots.}
    \label{reward3}
\end{figure}
Figure~\ref{dt33} evaluates the learning performance of the  DT-RPPO framework. In this framework, the DT model serves as a surrogate environment for agent training, enabling efficient policy development without direct interaction with the physical system. We compare the DT-RPPO with two schemes: the standard PPO algorithm with full knowledge of GU traffic and direct interaction with the real environment (denoted as Ideal PPO) and
PPO trained within the DT-SLAM framework without explicit uncertainty modeling (denoted as DT-PPO).

We first study the secure throughput with varying the interaction frequency between the DT and the real environment, as shwon in Fig~\ref{dt33}(a).
The x-axis represents the number of interactions between the UAVs and the real environment, where each information collection by the UAVs is counted as one interaction. The y-axis shows the achieved secure throughput, reflecting the effectiveness of the learned policy in the real environment.
Initially, both DT-based methods exhibit faster performance improvement compared to Ideal PPO, demonstrating the efficiency of training within a simulated environment. The DT-PPO and RPPO curves converge by approximately $4000$ interactions, whereas Ideal PPO requires around $5000$ interactions, indicating a $20\%$ reduction in convergence time due to the use of the DT. Ultimately, RPPO achieves performance on par with Ideal PPO, indicating that the DT accurately captures the dynamics of the real environment. In contrast, the growing performance gap between DT-PPO and RPPO during training reveals the accumulation of modeling errors in the absence of uncertainty awareness, thereby highlighting the critical importance of robust uncertainty management in simulation-based learning.
\begin{figure}
    \centering
    \subfloat[Mode-Switching.]{
\includegraphics[width=0.2\textwidth]{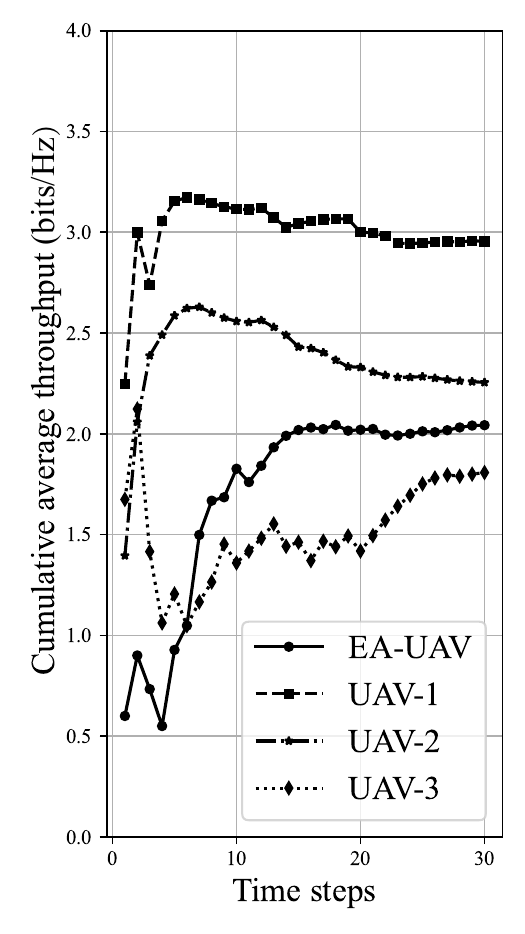}
    \label{fig:mode-switch-throughput}
    }
\hspace{-0.3cm}
    \subfloat[No-Jamming.]{
\includegraphics[width=0.2\textwidth]{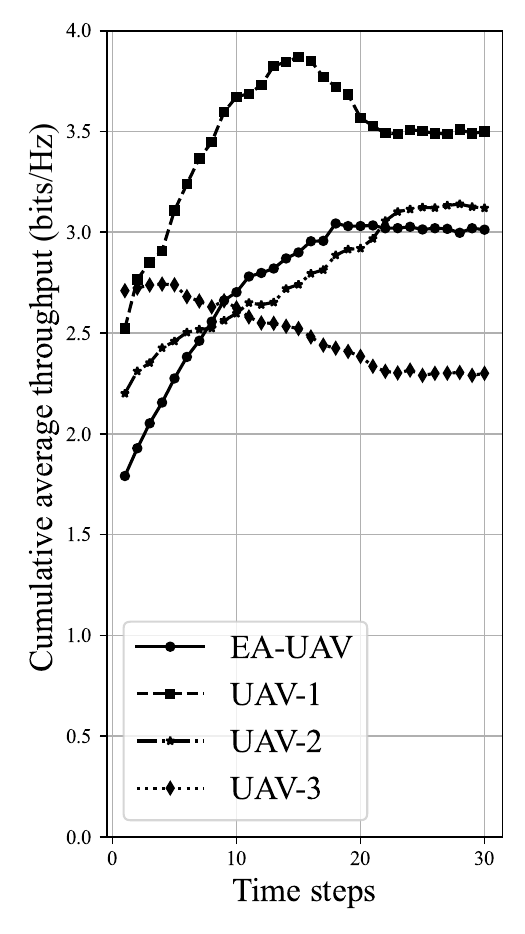}
    \label{fig:no-jamming-throughput}
    }
    \caption{UAVs' throughput under different jamming modes.}\label{throughput-all}
\end{figure}

Figure~\ref{dt33}(b) illustrates the training process of the LE-UAVs and the EA-UAV under a multi-stage Stackelberg game. The training alternates between optimizing the LE-UAVs’ secure communication strategies and updating the EA-UAV’s eavesdropping policy to improve interception performance. The red curve represents the EA-UAV’s training progress, and the same PPO algorithm is used for learning in all cases.
During LE-UAVs' training, improvements in trajectory planning, mode selection, and jamming coordination increase the overall rewards.
When the EA-UAV is trained, it enhances eavesdropping performance, temporarily degrading the overall reward.
This alternation continues until the reward stabilize, indicating convergence to a Stackelberg equilibrium.
Compared to Ideal PPO, RPPO achieves faster convergence and a smoother, more stable reward curve.
The digital twin enables efficient exploration with fewer real-world interactions, improving training stability.
The smaller reward drops during EA-UAV's retraining show that RPPO-trained LE-UAVs are more robust against eavesdropping, as the DT implicitly models adversarial behaviors, enabling proactive defense learning.
In contrast, DT-PPO exhibits notable oscillations. Limited real-environment feedback leads to insufficient exploration, resulting in suboptimal, overlapping UAVs' trajectories. This causes larger fitting errors in the GPR-based DT model, which increases simulation-to-reality gaps leading to lower secure throughput in real deployment.

\begin{figure*}[t]
	\centering
\subfloat[Mode-Switching.]{\includegraphics[width=0.322\textwidth]{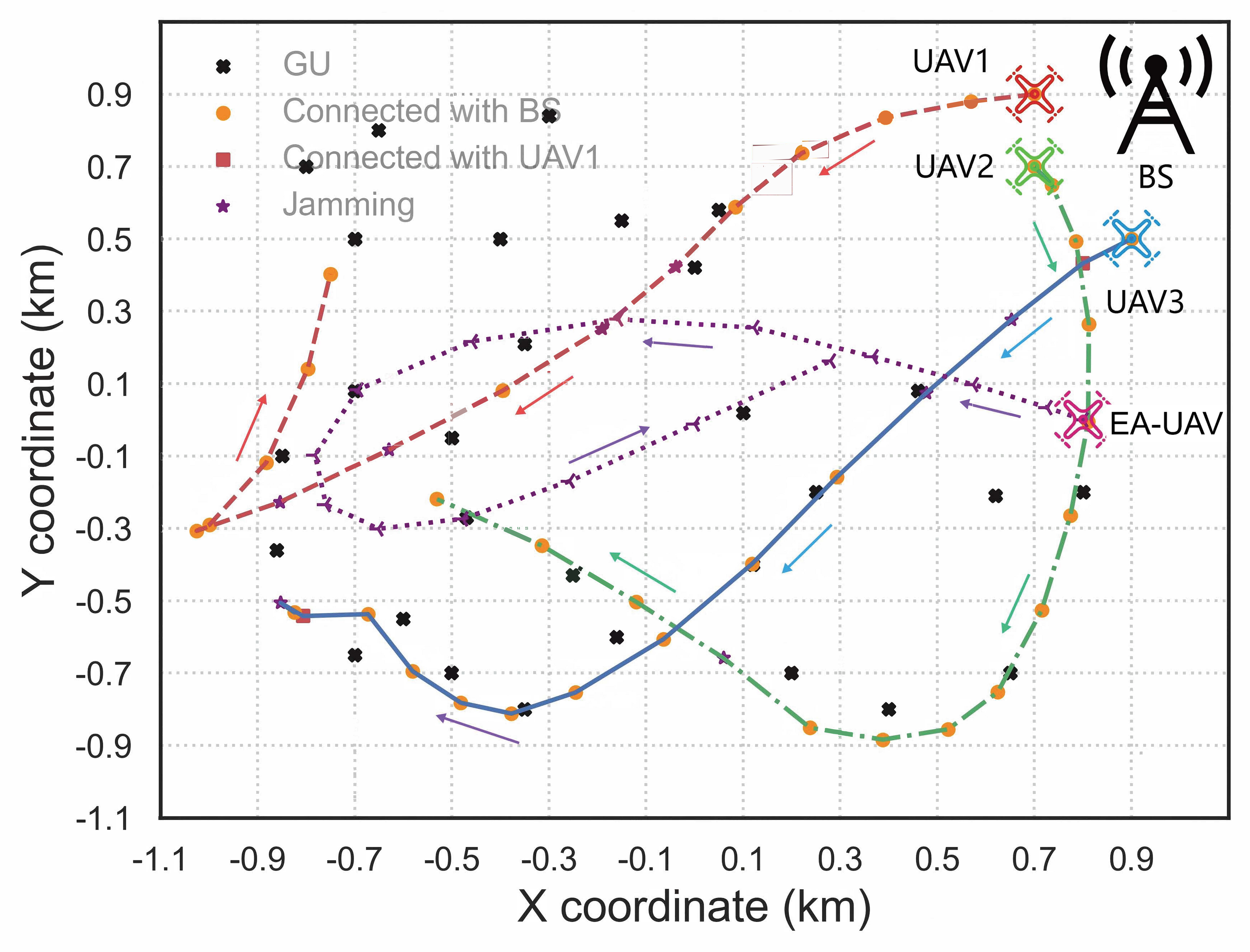}}
\subfloat[Fixed-Jamming.]{\includegraphics[width=0.33\textwidth]{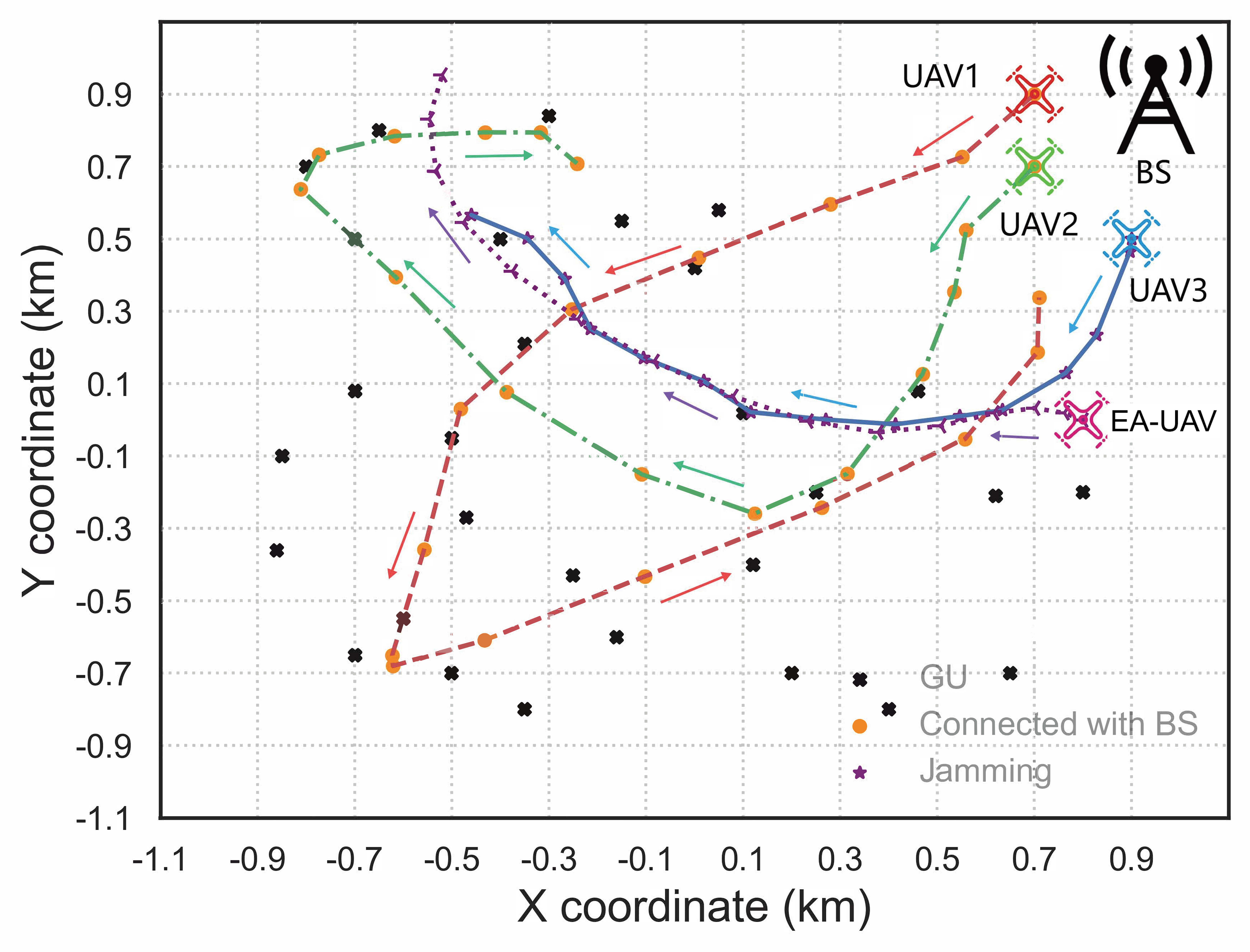}}
\subfloat[No-Jamming.]{\includegraphics[width=0.33\textwidth]{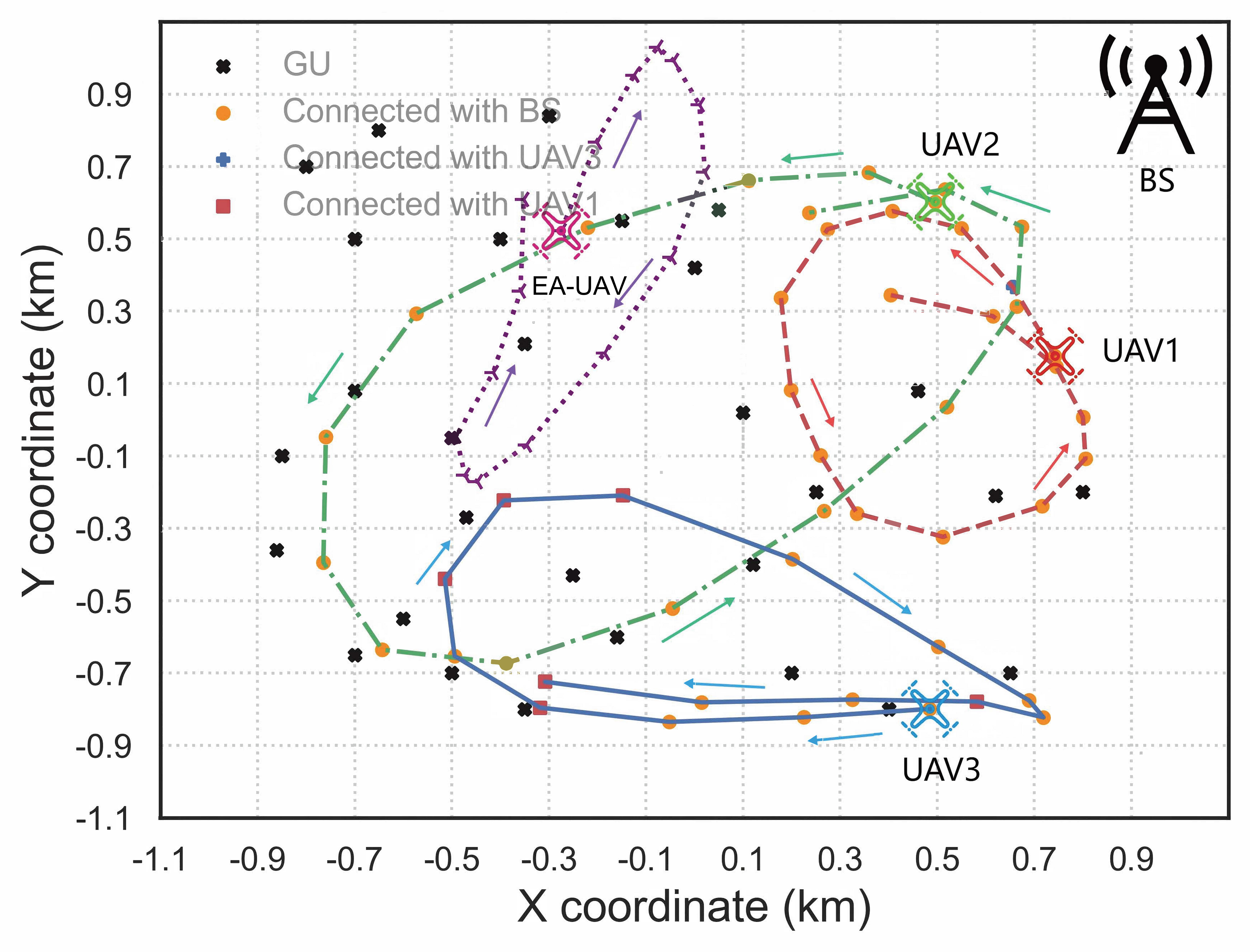}}
\caption{Trajectories with the different jamming strategies.}\label{tra}
\vspace{-0.4cm}
\end{figure*}

Figure~\ref{reward3} compares the cumulative average reward achieved by the DT-RPPO algorithm under different UAV jamming strategies.
The first strategy, referred to as Mode-Switching, corresponds to the proposed mechanism in which each UAV dynamically chooses between acting as a relay for data transmission or as a jammer to disrupt the eavesdropper.
The second strategy, Fixed-Jamming, assigns one UAV  exclusively to jamming while the other UAVs are dedicated to forwarding data from ground users.
The third case, No-Jamming, disables active jamming entirely, with UAVs attempting to avoid eavesdropping only through trajectory adjustments and scheduling. It is observed that the cumulative average reward of the jamming based schemes first increases and then converges to a stable value. This demonstrates that the RPPO algorithm is capable of obtaining stable data collection and trajectory policies for LE-UAVs. The No-Jamming scheme initially achieves high reward. However, without jamming against the EA-UAV, eavesdropping intensifies over time and ultimately degrades the overall reward.
It is also observed that the Fixed-Jamming and No-Jamming achieve comparable performance. This is because the fixed jamming assignment indiscriminately interferes with both sides. In contrast, the Mode-Switching scheme enables LE-UAVs to intelligently and flexibly allocate their roles based on real-time threat conditions.
As such, the Mode-Switching scheme highlights its superior effectiveness in enhancing secure throughput.

We further examine the cumulative average throughput of LE-UAVs and the EA-UAV under the Mode-Switching and No-Jamming schemes, as shown in Fig.~\ref{throughput-all}. For clarity, we denote the
$i$-th LE-UAV as UAV-$i$. Under the Mode-Switching scheme, the EA-UAV's throughput stabilizes at around 
$2$, which is lower than
$3$ achieved under the No-Jamming scheme. Although LE-UAVs experience a slight throughput reduction due to time spent in jamming mode, this trade-off effectively suppresses eavesdropping and improves overall secure communications. This validates the effectiveness of the proposed Mode-Switching strategy in countering eavesdropping threats.


\subsection{Trajectory Planning under Different Jamming Modes}

To investigate the impact of different jamming mechanisms on LE-UAVs’ secure transmission strategies, we present the trajectories and network topologies of multiple UAVs under three scenarios: Mode-Switching, Fixed-Jamming, and No-Jamming, all in the presence of the EA-UAV. Figure~\ref{tra}(a) illustrates the partial flight paths and communication topologies of the three UAVs when the RPPO algorithm converges in the presence of the EA-UAV. Under the Mode-Switching mechanism, the EA-UAV is unable to persistently follow any single LE-UAV for eavesdropping, as each LE-UAV can dynamically switch to jamming mode to disrupt its surveillance.
Initially, the EA-UAV approaches UAV-$3$ to intercept its transmissions. Upon detecting this threat, UAV-$3$ switches to active jamming, prompting the EA-UAV to shift its attention toward UAV-$1$. When UAV-$1$ also activates its jamming capability, the EA-UAV retreats to the center of the mission area. From this central position, it maintains proximity to all LE-UAVs and aligns itself along the potential relay paths between the LE-UAVs and the BS, attempting to maximize its eavesdropping opportunity.
Despite the persistent threat,  UAV-$1$, UAV-$2$, and UAV-$3$ achieve effective coordination. They exhibit a clear division of roles in data collection and relay, while taking turns to perform jamming. This dynamic cooperation ensures continuous suppression of the EA-UAV and significantly enhances the overall secure throughput.

Figure~\ref{tra}(b) depicts the UAVs' trajectories and network topology under the Fixed-Jamming mechanism. In this setup, UAV-$3$ is dedicated to jamming and continuously follows the EA-UAV while emitting interference signals. The other two LE-UAVs focus solely on data transmission and maintain direct communication links with the BS. UAV-$3$ successfully forces the EA-UAV toward the corner of the deployment area, limiting its ability to approach the data-relaying LE-UAVs.
Meanwhile, UAV-$1$ and UAV-$2$ establish stable connections with the BS to ensure reliable data delivery. Due to the reduced number of active relays, each transmitting UAV serves a larger subset of ground users, leading to increased individual data loads. As a result, maintaining direct UAV-BS links becomes the most efficient strategy, minimizing transmission delay and energy consumption. This highlights the trade-off between dedicated jamming and relay efficiency in resource-constrained multi-UAV networks.

Figure~\ref{tra}(c) presents the scenario without jamming. In this case, the EA-UAV dynamically adjusts its trajectory in real time based on the positions and movements of the LE-UAVs. The multi-UAV system lacks accurate knowledge of the EA-UAV’s strategy and therefore cannot fully evade its surveillance. Over time, a game-theoretic equilibrium is established between the LE-UAVs and the EA-UAV.
Regarding cooperation among the LE-UAVs, UAV-$1$ operates near the BS and is responsible for local data collection. It also serves as a relay to forward data from UAV-$2$ and UAV-$3$ to the BS. UAV-$2$ follows a wide loop to gather data from distant regions and transmits when the channel condition to the BS is favorable. UAV-$3$ focuses on coverage in areas far from the BS and relies on UAV-$1$ as an intermediate relay for data delivery.
When both sides reach equilibrium, their trajectories stabilize. At this point, the EA-UAV succeeds in intercepting a significant amount of transmitted data, while the LE-UAVs continue to deliver information under the best possible security conditions given the absence of active countermeasures.

\subsection{Secure Throughput of RPPO Framework}

\begin{figure}
    \centering
    \subfloat[Different user size.]{
\includegraphics[width=0.2\textwidth]{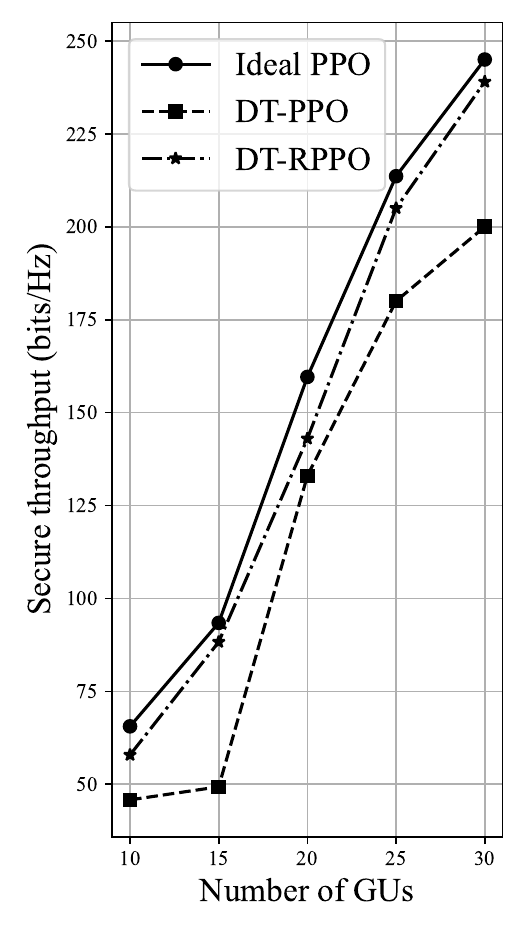}
    \label{dt}
    }
\hspace{-0.3cm}
    \subfloat[Different jamming power.]{
\includegraphics[width=0.2\textwidth]{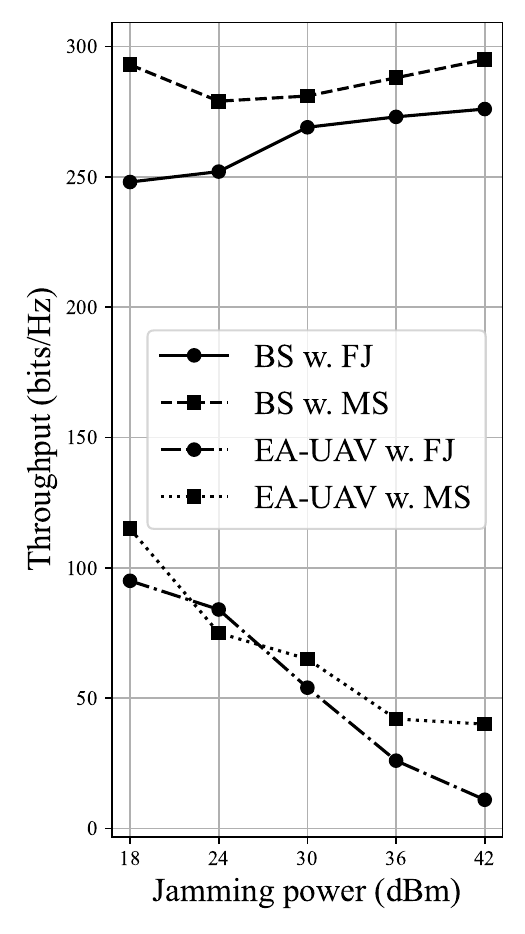}
    \label{jam}
    }
    \caption{ Throughput performance under different schemes.}\label{alll-1}
\end{figure}

Figure~\ref{alll-1}(a) shows the secure throughput of Ideal PPO, DT-PPO, and DT-RPPO under varying numbers of GUs. Ideal PPO, trained directly in the real environment, serves as the performance upper bound.
Secure throughput increases with the number of GUs due to higher data volume. When GUs are few, their sparse distribution makes individual locations critical for UAV routing. In this regime, DT-PPO exhibits suboptimal performance due to insufficient exploration control, resulting in inaccurate modeling of individual GU locations within the digital twin. This leads to a noticeable gap in secure throughput compared to Ideal PPO.
As the GU count reaches $20$, the environment becomes denser, improving the statistical reliability of the DT model.
DT-PPO then achieves near-optimal performance with minimal degradation.  However, beyond this point, environmental complexity grows rapidly. Without explicit uncertainty modeling, DT-PPO accumulates observation errors and suffers reduced model fidelity, degrading policy performance.
In contrast, RPPO consistently maintains throughput close to Ideal PPO across all scenarios. Its resilience stems from uncertainty-aware reinforcement learning, which enhances exploration and improves DT robustness. This demonstrates that RPPO is not only effective but also adaptable to changes in network scale and environmental complexity.

Figure~\ref{alll-1}(b) shows the throughput of the BS and the EA-UAV under Fixed-Jamming (denoted as FJ) and Mode-Switching (denoted as MS) schemes.
In the No-Jamming case, no interference is introduced. As such, both the BS and the EA-UAV achieve the maximum throughput. 
Under Fixed-Jamming, EA-UAV throughput decreases steadily as jamming power increases due to stronger signal suppression. BS throughput first increases thanks to improved security and then saturates when jamming becomes excessive. 
For Mode-Switching, at low jamming power (e.g., $18$ dBm), UAVs prefer transmission mode to avoid sacrificing communication opportunities. As jamming power increases, each jamming action becomes more effective, prompting UAVs to switch to jamming mode more frequently to enhance security but reducing BS throughput due to less transmission time. At high jamming power, even brief jamming is sufficient to deter the eavesdropper, allowing UAVs to return to transmission mode and recover BS throughput.

\begin{figure}[t]
    \centering
    \subfloat[Runtime.]{
\includegraphics[width=0.2\textwidth]{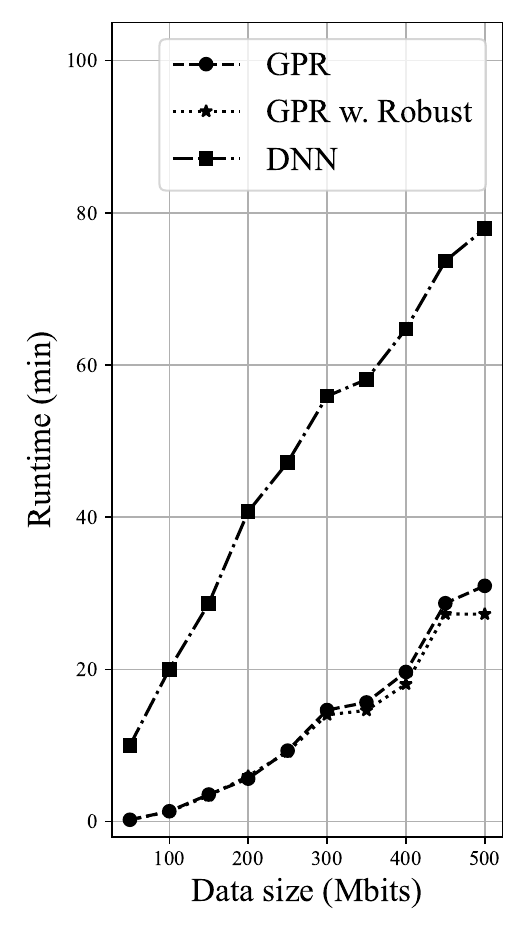}
    \label{dt}
    }
\hspace{-0.3cm}
    \subfloat[Model mismatch.]{
\includegraphics[width=0.23\textwidth]{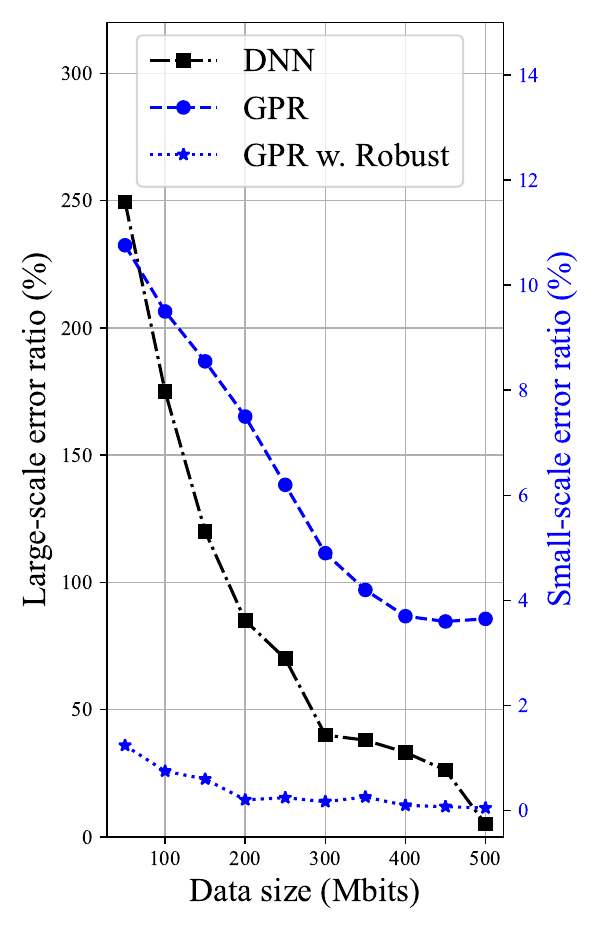}
    \label{jam}
    }
    \caption{ DT evaluation under different schemes.}\label{dt3}
\end{figure}

Figure~\ref{dt3} compares the computational time and modeling accuracy of different algorithms used to construct the DT. The three methods evaluated are the uncertainty-aware GPR (denoted as GPR w. Robust), standard GPR, and a deep neural network (DNN) approach. In Fig.~\ref{dt3}(a), the horizontal axis represents the number of interaction samples collected by the UAV from the real environment, denoted as $|\mathcal{M}|$. The vertical axis shows the time required to build the DT model in minutes.
It can be observed that the DNN method requires significantly more time than both GPR-based approaches. This is due to the higher computational complexity of DNNs, which involve a large number of parameters and iterative optimization during training. In contrast, GPR methods achieve faster model construction owing to their analytical formulation and lower training overhead.
The GPR w. Robust and the standard GPR exhibit similar computation times in the early stages of data collection. However, as the number of samples increases, the GPR w. Robust achieves slightly lower model construction time. This improvement stems from the structure of the kernel matrix $k(X,X)$, which must be inverted during the GPR training. When data points are clustered or highly correlated, the kernel matrix becomes ill-conditioned, leading to numerical instability and longer inversion time. By promoting more extensive exploration of the task area, GPR w. Robust yields better spatial distribution of collected data. This results in a more stable and well-conditioned kernel matrix, reducing the computational burden of matrix inversion and improving overall efficiency.

Figure~\ref{dt3}(b) illustrates the relationship between the number of collected data points and the accuracy of the DT. The black line corresponds to the left y-axis, representing modeling error in percentage, while the blue line aligns with the right y-axis, indicating the relative accuracy level.
To evaluate DT accuracy, we adopt a relative error metric computed from uniformly sampled points across the task area. For each point $n$, the deviation between the fitted value $o_n$ and the true observation $s_n$ is computed. The average relative error is then calculated as $\frac{1}{N}\sum_{n}^N\frac{\sqrt{||s_n-o_n||^2}}{s_n} \times 100\%$.
The results show that under limited data, the DNN exhibits significantly higher error compared to both GPR-based methods. This is because GPR leverages prior knowledge encoded in the kernel function, enabling effective interpolation and robust performance even with sparse observations. As the amount of training data increases, the DNN gradually improves and its error converges to approximately $1\%$, reflecting its strong fitting capability in data-rich conditions.
In contrast, both standard GPR and the proposed GPR w. Robust achieve low error from the early stages of learning, demonstrating their sample efficiency. Notably, GPR w. Robust further reduces error by employing uncertainty-guided exploration, which directs the UAV to acquire data from regions of high information gain. This leads to faster model convergence and a more accurate representation of the real environment, highlighting the advantage of active learning through uncertainty control in digital twin construction.

\section{Conclusion}\label{7}
This paper has investigated a multi-UAV secure communication system under intelligent eavesdropping threats.  We have proposed an intelligent mode-switching mechanism that enables UAVs to dynamically switch between transmission and jamming modes, improving both communication efficiency and security.
We have formulated a joint optimization problem of UAVs' trajectories, network formation, and model selection, as well as the GUs' transmission control to maximize the system’s secure throughput. To capture interactions between the LE-UAVs and the EA-UAV, we have modeled the problem as a multi-stage Stackelberg game and solve it by alternately optimizing UAV strategies and eavesdropper responses.
We have designed a DT-SLAM framework, where the DT acts as a virtual replica of the physical environment. This framework reduces the reliance on real-world interactions by providing synthetic training data for DRL, significantly improving learning efficiency. Furthermore,  we have proposed the RPPO algorithm, which evaluates the model mismatch between the real and the DT environment and integrates it into the DRL learning. This enables robust policy learning by accounting for discrepancies between the DT and the real world.
Simulations show that DT-RPPO achieves better stability and faster convergence than Ideal PPO. Additionally, the mode-switching scheme  achieves higher throughput and better security in highly dynamic network environments.
\bibliographystyle{IEEEtran}
\bibliography{sample}
\end{document}